\documentclass[aip,reprint]{revtex4-2}
\usepackage{amsmath}
\usepackage{mathtools}
\usepackage{amssymb}
\usepackage{xcolor}

\usepackage{graphicx}
\usepackage{bm}

\begin{document}


\title{Phonon Chirality Induced by Vibronic-Orbital Coupling}
\author{Yun-Yi Pai}
\email{yunyip@ornl.gov}
\address{Materials Science and Technology Division, Oak Ridge National Laboratory, Oak Ridge, TN 37831, USA}
\address{Quantum Science Center, Oak Ridge, Tennessee 37831, USA}

\author{Claire E. Marvinney}
\address{Materials Science and Technology Division, Oak Ridge National Laboratory, Oak Ridge, TN 37831, USA}
\address{Quantum Science Center, Oak Ridge, Tennessee 37831, USA}

\author{Liangbo Liang}
\address{Center for Nanophase Materials Sciences, Oak Ridge National Laboratory, Oak Ridge, TN 37831, USA}

\author{Ganesh Pokharel}
\address{Materials Department and California Nanosystems Institute, University of California Santa Barbara, Santa Barbara, CA 93106, USA}

\author{Jie Xing}
\address{Materials Science and Technology Division, Oak Ridge National Laboratory, Oak Ridge, TN 37831, USA}

\author{Haoxiang Li}
\address{Materials Science and Technology Division, Oak Ridge National Laboratory, Oak Ridge, TN 37831, USA}

\author{Xun Li}
\address{Materials Science and Technology Division, Oak Ridge National Laboratory, Oak Ridge, TN 37831, USA}

\author{Michael Chilcote}
\address{Materials Science and Technology Division, Oak Ridge National Laboratory, Oak Ridge, TN 37831, USA}
\address{Quantum Science Center, Oak Ridge, Tennessee 37831, USA}

\author{Matthew Brahlek}
\address{Materials Science and Technology Division, Oak Ridge National Laboratory, Oak Ridge, TN 37831, USA}
\address{Quantum Science Center, Oak Ridge, Tennessee 37831, USA}

\author{Lucas Lindsay}
\address{Materials Science and Technology Division, Oak Ridge National Laboratory, Oak Ridge, TN 37831, USA}

\author{Hu Miao}
\address{Materials Science and Technology Division, Oak Ridge National Laboratory, Oak Ridge, TN 37831, USA}

\author{Athena S. Sefat}
\address{Materials Science and Technology Division, Oak Ridge National Laboratory, Oak Ridge, TN 37831, USA}

\author{David Parker}
\address{Materials Science and Technology Division, Oak Ridge National Laboratory, Oak Ridge, TN 37831, USA}

\author{Stephen D. Wilson}
\address{Materials Department and California Nanosystems Institute, University of California Santa Barbara, Santa Barbara, CA 93106, USA}

\author{Benjamin J. Lawrie}
\email{lawriebj@ornl.gov; This manuscript has been authored by UT-Battelle, LLC, under contract DE-AC05-00OR22725 with the US Department of Energy (DOE). The US government retains and the publisher, by accepting the article for publication, acknowledges that the US government retains a nonexclusive, paid-up, irrevocable, worldwide license to publish or reproduce the published form of this manuscript, or allow others to do so, for US government purposes. DOE will provide public access to these results of federally sponsored research in accordance with the DOE Public Access Plan (http://energy.gov/downloads/doe-public-access-plan). }
\address{Materials Science and Technology Division, Oak Ridge National Laboratory, Oak Ridge, TN 37831, USA}
\address{Quantum Science Center, Oak Ridge, Tennessee 37831, USA}

\date{\today}

\begin{abstract}
The notion that phonons can carry pseudo-angular momentum has become popular in the last decade, with recent research efforts highlighting phonon chirality, Berry curvature of phonon band structure, and the phonon Hall effect.  When a phonon is resonantly coupled to a crystal electric field excitation, a so-called vibronic bound state forms. Here, we observe angular momentum transfer of $\Delta \textbf{J}_z = \pm 1 \hbar$ between phonons and an orbital state in a vibronic bound state of a candidate quantum spin liquid. This observation has profound implications for the engineering of phonon band structure topology through chiral quasiparticle interactions.

\end{abstract}


\maketitle

\section{Introduction}
Since the seminal work of Bloch in 1929\cite{Bloch1929}, electron-phonon coupling has emerged as one of the most important topics in modern condensed matter physics. In conventional superconductors, electron-phonon coupling underlies Cooper pairing.  In semiconductors, it sets an upper bound for electron mobility.  Electron-phonon coupling also governs numerous thermal and spin relaxation processes in solids. In ionic crystals with f orbitals, spin-orbit coupling and the crystal electric field (CEF) associated with the ionic or ligand environment split the electronic wavefunction eigenstates into manifolds. Atomic displacements can change the ligand environments and hence the orbital manifolds. This change is typically treated within the adiabatic Born-Oppenheimer approximation, in which atomic motion is static relative to the electronic degrees of freedom. However, when phonon-CEF coupling is nearly resonant, the Born-Oppenheimer approximation is inadequate and a new state -- a vibronic bound state (VBS), with both orbital character and phonon character -- can form.  

Phonons themselves can have chirality and carry pseudo angular momentum\cite{chiralphonon_2D:review, PhysRevLett.115.115502, JUNEJA2021100548, PhysRevB.98.241405, PhysRevMaterials.5.085002, coh2021classification} when inversion symmetry~($\mathcal{P}$) or time reversal symmetry~($\mathcal{T}$) is broken\cite{chiralphonon, Chiral_CrBr3}.  The ability to control pseudo angular momentum transfer between phonons and electronic degrees of freedom will unlock new opportunities in information processing. However, the transfer of angular momentum between photons, electronic spin, orbital excitations, and phonons is still not well understood. It was studied, for instance, in the context of paramagnetic spin relaxation\cite{manenkov1966spin, deHaas2}, but phonons are not generally thought to carry macroscopic angular momentum\cite{deHaas2, PAM}. Phononic pseudo angular momentum is key to several fundamental effects in physics, including the microscopic explanation of the phonon Hall effect\cite{PhononHall}, Einstein-de Haas effect \cite{EinsteinDeHaas}, possible circular heat flow\cite{circularHeat}, and zero point energies of chiral phonons \cite{EinsteinDeHaas}. The transfer channel from spin to phonon, on the other hand, is thought to be relevant in ultra-fast demagnetization processes \cite{Chen19258, ultraDemag}.

Here, we report a change of angular momentum $\Delta \textbf{J}_z = \pm 1 \hbar$ in an orbital excitation due to phononic coupling in NaYbSe$_2$.  NaYbSe$_2$\cite{MottSc_NaYbSe2,NaYbSe2_PRX,NaYbSe2_2,NaYbSe2PRB_CEF,zhang2020pressure} belongs to the family of Yb delafossites with the form A$^{1+}$Yb$^{3+}$X$^{2-}_{2}$ that have been identified as quantum spin liquid (QSL) candidates. The QSL is a theoretically proposed phase that is characterized by massive many-body entanglement\cite{ANDERSON1973153,Knolle2019,Savary2016} with possible application to fault-tolerant quantum information processing\cite{barkeshli2014coherent}.  So far 14 members within this family have been identified to date\cite{Schmidt_2021,NaYbS2_2018,NaYbS2_2019,NaYbSe2PRB_CEF,NaYbSe2_2,NaYbSe2_PRX,NaYbO2_1, NaYbO2_3, NaYbO2_SDW,Xing_Jie_PRB2019Rapid,pocs2021systematic,xie2021fieldinduced,RN3955, pai2021nearlyresonant,xing2021synthesis,scheie2021witnessing}. They all have planar triangular lattices with spin S$_{\text{eff}} = 1/2$ on trigonally distorted YbX$_6$ octahedra. Antiferromagnetic coupling in the triangular lattice causes geometric frustration, resulting in a lack of long-range magnetic order down to the lowest probed temperatures. This family of candidate QSLs is objectively less defect prone than Yb(Mg, Ga)O$_4$\cite{NaYbSe2PRB_CEF}, and the breadth of substitutional composites in the family makes it an ideal platform for the study and control of QSL excitations.

The effective spin S$_{\text{eff}}$ = 1/2 of the system comes from the Yb$^{3+}$ ion, which has the [Xe]4f$^{13}$ electronic configuration. Yb$^{3+}$, like all the elements in the 4f block, has weak exchange coupling and strong spin-orbit coupling compared to the 3d-block elements.  The ground state spin-orbit manifold of Yb$^{3+}$ has total spin $J = 7/2$. The next manifold $J = 5/2$ is more than an eV above the ground state manifold\cite{Koningstein_1967, Petit}.  The ground-state $J = 7/2$ manifold splits into 4 time-reversed Kramers pairs. This degeneracy is a result of Kramers theorem: trivalent Yb$^{3+}$ has an odd number of electrons and a  $\mathcal{T}$-symmetric Hamiltonian, so the eigenenergies are doubly degenerate at zero magnetic field.  The CEF splitting within the ground-state manifold is typically only tens of meV due to the well-shielded nature of the 4f electrons. 

VBSs form when strong CEF-phonon coupling is present, but they have only been reported in a handful of materials\cite{k6695,Sethi_PRL_2019_omegas,JTCeAl2, RamanCeAl2, Thalmeier_1982,adroja2012vibron,CEFPhononPrNi5,CEF_highTC,Gaudet_CEF_ph_Ho2Ti2O7,CEFEatenConfigCrossOver}. Structurally, NaYbSe$_2$ has less distortion in its YbSe$_6$ octahedra than other Yb delafossites\cite{Schmidt_2021}. Its ground states are reported to be described by spinon excitations \cite{NaYbSe2_PRX} or ferrimagnetic quasistatic and dynamic excitations within a QSL matrix\cite{zhu2021fluctuating}. While pristine NaYbSe$_2$ is insulating, with increasing pressure it exhibits increased conductivity and eventually demonstrates a dip in resistivity attributed to a possible superconducting state\cite{MottSc_NaYbSe2, zhang2020pressure}.  

\section{Results and Discussion}
\begin{figure*}
\centering
    \includegraphics[width=2\columnwidth]{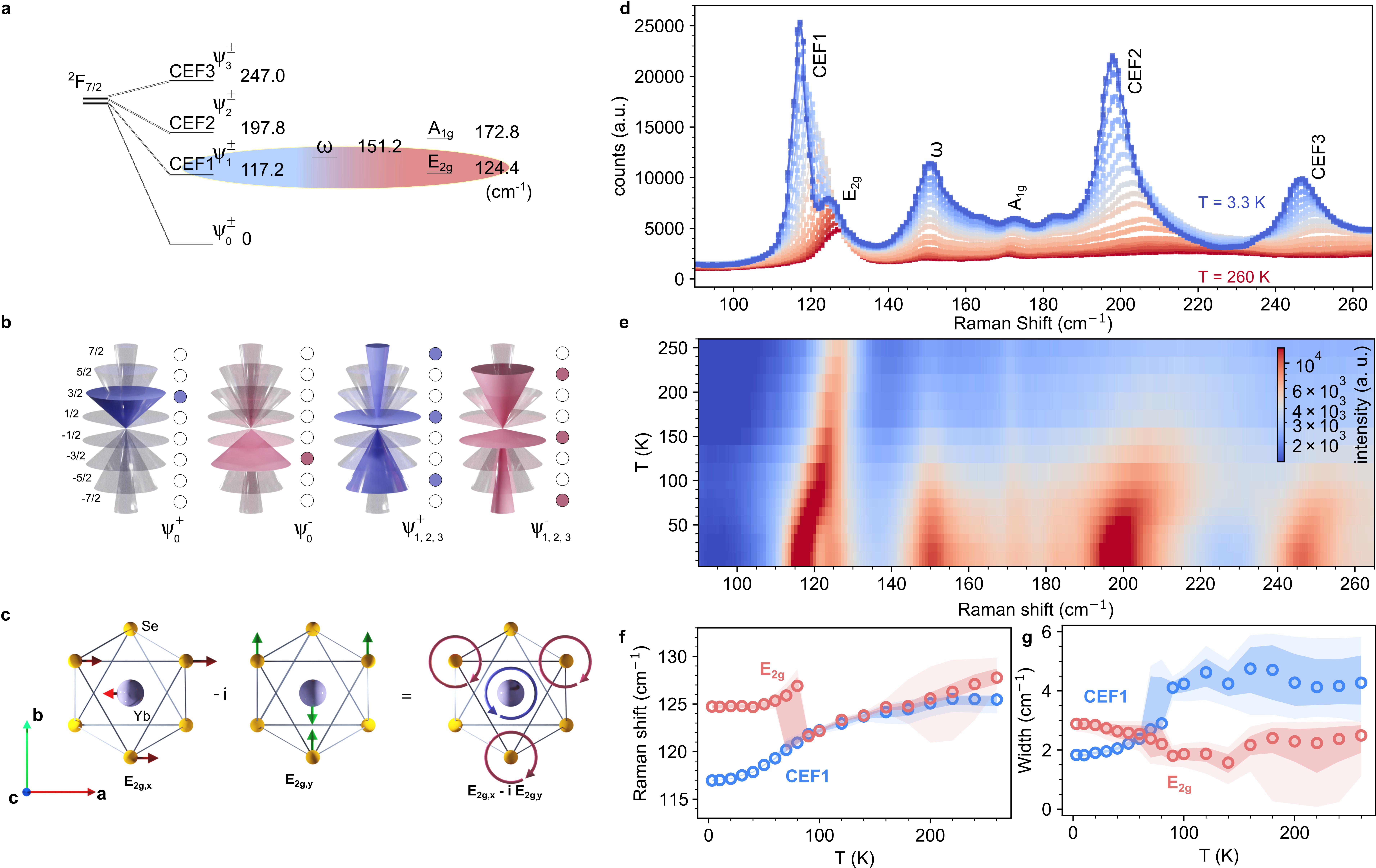}
    \caption{(a) Energy diagram of the ground state $J = 7/2$ manifold CEF levels, Raman active phonon modes, and vibronic bound state $\omega$. (b) A semiclassical representations of the CEF eigenstates $| \psi^{\pm}_0 \rangle$ and $| \psi^{\pm}_{1, 2, 3} \rangle$. The eigenstates are linear combinations of multiplets $m_J = -7/2 \dots 7/2$.  For example, $| \psi^+_0 \rangle =  | \frac{7}{2}, \frac{3}{2} \rangle $ is represented by precessing cones with only $c_{\frac{3}{2}} \neq 0$ for $ | \psi^+_0 \rangle = c_{\frac{3}{2}} |J,  \frac{3}{2} \rangle$. Hence all but the $ m_J  = \frac{3}{2}$ cone are transparent (and all but the $ m_J  = \frac{3}{2}$ circle are empty). The same follows for $| \psi^{-}_0 \rangle$ and $| \psi^{\pm}_{1, 2, 3} \rangle$. (c) The  E$_{\text{2g}}$ eigenspace is spanned by the basis \{$\text{E}_{\text{2g, x}}$, $\text{E}_{\text{2g, y}}$\}, which can also be spanned by \{$\text{E}_{\text{2g,+}}$, $\text{E}_{\text{2g,-}}$\} = \{$\text{E}_{\text{2g, x}} + i\text{E}_{\text{2g, y}}$, $\text{E}_{\text{2g, x}} - i\text{E}_{\text{2g, y}} $\}. (d) Temperature dependent Raman spectra from $T = 3.3$ K to $T = 260$ K.  (e) The contour plot of the temperature dependent data from (d). (f), (g) Peak positions and widths for E$_{\text{2g}}$ and CEF1 from Bayesian inference. The symbols are medians of the posterior distribution, and the 68\% and 95\% highest density intervals (HDIs) are represented by darker and lighter shading, respectively.}
    \label{fig:temp}
\end{figure*}

Figure \ref{fig:temp}a shows the low energy excitations of NaYbSe$_2$, where the ground state spin-orbit manifold $J = 7/2$ splits into 4 time-reversed Kramers pairs $|\psi_0^\pm\rangle$, $|\psi_1^\pm\rangle$, $|\psi_2^\pm\rangle$, $|\psi_3^\pm\rangle$.  CEF1, CEF2, and CEF3 describe the transition between $|\psi_1^\pm\rangle$, $|\psi_2^\pm\rangle$, $|\psi_3^\pm\rangle$ and the ground state $|\psi_0^\pm\rangle$, respectively. The CEF modes are observed at 117.2 cm$^{-1}$ (CEF1), 197.8 cm$^{-1}$ (CEF2), and 247.0 cm$^{-1}$ (CEF3) in Raman spectra acquired at 3.3 K, consistent with previous reports\cite{NaYbSe2PRB_CEF,pai2021nearlyresonant}. NaYbSe$_2$ also has two Raman-active phonon modes: E$_{\text{2g}}$ at 124.4 cm$^{-1}$ and A$_{\text{1g}}$ at 172.8 cm$^{-1}$.  Figure \ref{fig:temp}b shows a semiclassical \textit{spinning top} representation of the spin-orbit eigenstates for $|\psi_0^\pm\rangle$, $|\psi_1^\pm\rangle$, $|\psi_2^\pm\rangle$, $|\psi_3^\pm\rangle$ as $\sum_{m_j = -\frac{7}{2}, ...\frac{7}{2}} c_{m_j} |\frac{7}{2}, m_j \rangle$. Transparent cones represent $c_{m_j} = 0$. The blue (pink) cones represent the $+$ ($-$) branch.  

\begin{figure*}[t]
\centering
    \includegraphics[width=2\columnwidth]{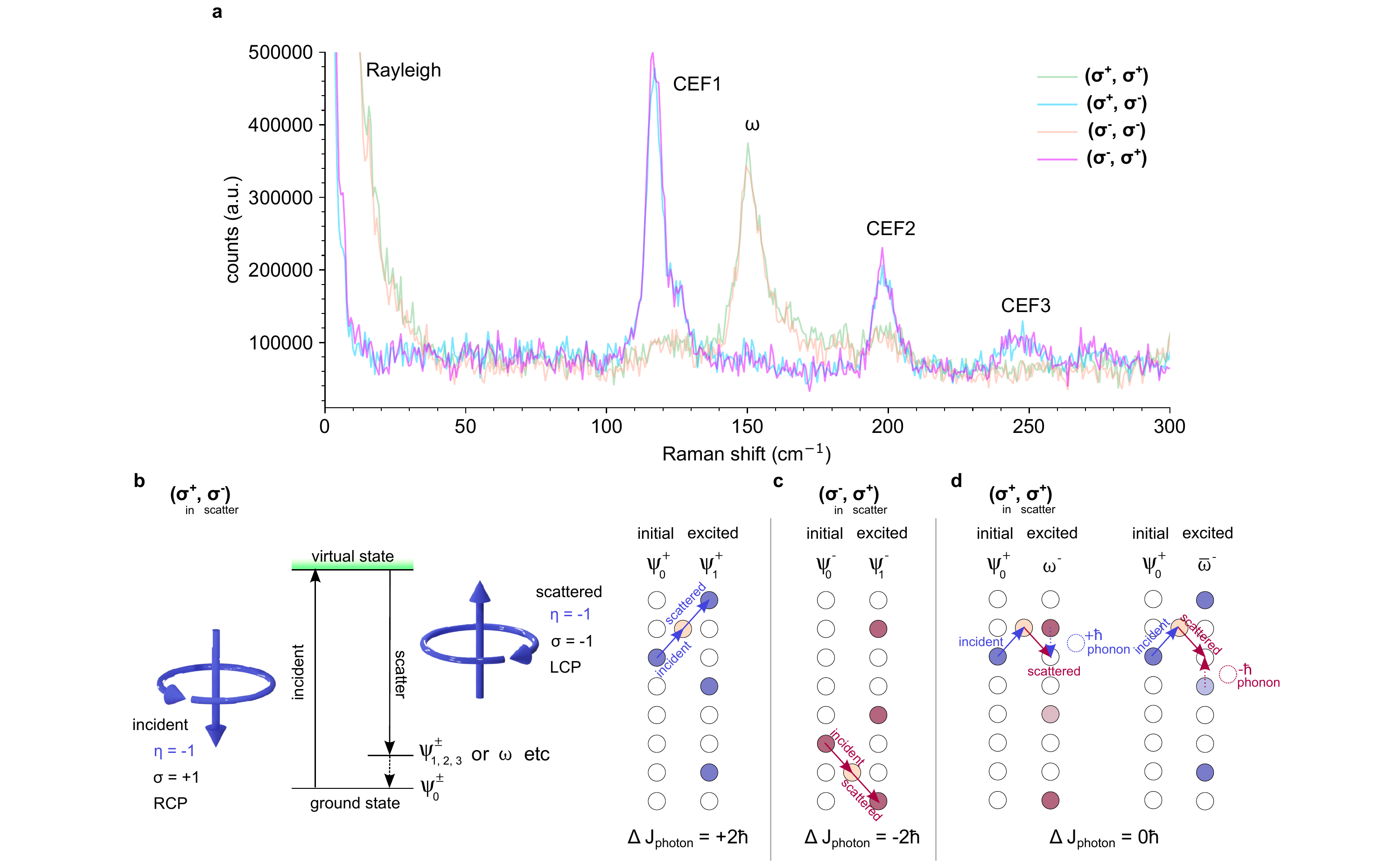}
    \caption{(a) Helicity resolved Raman spectra at $T = 4 $ K, $B = 0$ T. Four prominent modes are CEF1 (with E$_{\text{2g}}$ at its shoulder), $\omega$, CEF2, and CEF3. The Rayleigh scattering and $\omega$ are of ($\sigma^+$, $\sigma^+$) and ($\sigma^-$, $\sigma^-$) co-circular scattering channels, while CEF1-3 are of ($\sigma^+$, $\sigma^-$) and ($\sigma^-$, $\sigma^+$) cross-circular channels. (b-d) Selection rules for CEF modes and $\omega$ with virtual states illustrated in light pink.}
    \label{fig:eigen_sele}
\end{figure*}

In this material, the doubly degenerate E$_{\text{2g}}$ mode is resonant with the CEF1 state. This resonance gives rise to a VBS, $\omega$ (Figure \ref{fig:temp}a). As illustrated in Figure \ref{fig:temp}c, the E$_{\text{2g}}$ mode is spanned by the basis \{$\text{E}_{\text{2g, x}}$, $\text{E}_{\text{2g, y}}$\}; it can also be spanned by the chiral basis \{$\text{E}_{\text{2g,+}}$, $\text{E}_{\text{2g,-}}$\} = \{$\text{E}_{\text{2g, x}} + i\text{E}_{\text{2g, y}}$, $\text{E}_{\text{2g, x}} - i\text{E}_{\text{2g, y}} $\}\cite{helicity_Umklapp}. This second  basis is the natural one for CrBr$_3$ because of broken $\mathcal{T}$-symmetry\cite{Chiral_CrBr3}. In NaYbSe$_2$, however, because both $\mathcal{P}$ and $\mathcal{T}$ are intact, phonons are not intrinsically chiral, and the chiral basis is not, at first glance, the appropriate choice of basis. 

Using temperature dependent Raman spectra, we first verify the primary CEF excitations and phonon modes. The spectra are  shown in Figures \ref{fig:temp}d (traces) and \ref{fig:temp}e (contour).  The CEF modes are clearly identified at 117.2 cm$^{-1}$ (CEF1), 197.8 cm$^{-1}$ (CEF2), and 247.0 cm$^{-1}$ (CEF3) in the spectra. Additionally, they all become significantly stronger in intensity, and soften (shift toward lower energy) as the temperature decreases. On the other hand, the temperature dependence of the E$_{\text{2g}}$ and the A$_{\text{1g}}$ phonon modes is far less drastic. These CEF and phonon modes are consistent with recently reported linearly polarized NaYbSe$_2$ Raman spectra\cite{NaYbSe2PRB_CEF}. Note that the E$_{\text{2g}}$ and CEF1 modes are closely spaced. This is the case for NaYbSe$_2$\cite{NaYbSe2PRB_CEF}, CsYbSe$_2$\cite{pai2021nearlyresonant}, and KYbSe$_2$\cite{scheie2021witnessing}. Using Bayesian inference with a Hamiltonian Monte Carlo \texttt{PyMC3}\cite{pyMC}, we track the peak parameters from data in Figure \ref{fig:temp}d. The extracted peaks and widths for CEF1 and E$_{\text{2g}}$ are shown in Figure \ref{fig:temp}f and \ref{fig:temp}g, respectively. The symbols represent the median values from the Bayesian inference. The two shaded bands illustrate the 68\% (darker) (corresponding to 1$\sigma$ in the central limit) and 95\% (lighter) (corresponding to 2$\sigma$) highest density intervals (HDIs) (See Supplementary Information for other selected modes).  The model has large error bars from $T = 175$ K to $225$ K, but the two modes are well resolved elsewhere. The rapid change in the E$_{\text{2g}}$ Raman mode energy at 90 K and the associated overlap of this mode with the CEF1 mode from 100 to 200 K is noteworthy. In CsYbSe$_2$, there is some evidence for anisotropic negative thermal expansion in these temperature ranges\cite{Deng:wm6047, Xing_Jie_PRB2019Rapid}. In general, a nearly discontinuous change with temperature of any spectral parameter (such as the energy of the E$_{\text{2g}}$ mode) is suggestive of a phase transition, though we find no direct evidence of any phase transition here. 

At the same time, we see the VBS mode $\omega$ at 151.2 cm$^{-1}$. It has characteristics in common with both CEFs and phonons: $\omega$ becomes significantly stronger at low temperatures just as the CEFs do, and it hardens as the temperature decreases like A$_{\text{1g}}$ does. The mode $\omega$ is present in both NaYbSe$_2$ and CsYbSe$_2$, although it is far less prominent in CsYbSe$_2$. This is likely because the spacing between CEF1 and E$_{\text{2g}}$ is smaller in NaYbSe$_2$ (7.2 cm$^{-1}$) than in CsYbSe$_2$ (12.9 cm$^{-1}$)\cite{pai2021nearlyresonant}. Fitting to the Thalmeier-Fulde description of a magnetoelastic vibronic bound state\cite{Thalmeier_1982} yields a coupling strength of 32.0 cm$^{-1}$ (3.97 meV) for NaYbSe$_2$, which is far stronger than the 23.6 cm$^{-1}$  (2.93 meV) reported for CsYbSe$_2$ and comparable to the roughly 34.0 cm$^{-1}$  (4.22 meV) coupling strength reported for Ce$_2$O$_3$\cite{Sethi_PRL_2019_omegas}. 

Typically, CEF spectroscopy data are fit to a CEF Hamiltonian with Stevens operators given by the symmetry of the ionic environment of the 4f orbitals (Methods). This links the CEF peaks to eigenstates $| \psi \rangle$. The procedure is typically under-constrained, but that is not the case when optical selection rules are taken into account.  Figure \ref{fig:eigen_sele}a illustrates helicity-resolved Raman spectra acquired at $T=4$ K, with CEF1, CEF2, CEF3 and $\omega$ highlighted. While Rayleigh scattering is stronger in the co-circular polarization configuration, CEF1--CEF3 are stronger in the cross-circular polarization configuration. The mode $\omega$ is also stronger in the co-circular polarization configuration. 

Each right (left) circularly polarized $\sigma^+$ ($\sigma^-$) photon carries angular momentum $+\hbar$ ($-\hbar$). The link between helicity and circular polarization is given by $h$~=~\textbf{$\sigma$}~$\cdot$~$\textbf{k}$, where $\textbf{k}$ is the momentum of the photon, and $h$ the helicity\cite{Helicity}. Absorption of a $\sigma^+$ photon increases the angular momentum of the lattice by $+\hbar$, corresponding to being acted on by operator $J_+$, while scattering a $\sigma^+$ photon corresponds to $J_-$ acting on the lattice. Therefore, if one considers Raman spectra acquired in the ($\sigma_{\text{incident}}$, $\sigma_{\text{scatter}}$) = ($\sigma^+$, $\sigma^-$) configuration, as in Figure \ref{fig:eigen_sele}b, the system obtains $+2\hbar$ of angular momentum, corresponding to $J_+ J_+$ acting on it (a generalization of the closure condition in Axe et al. \cite{Closure}, which contracts $\sum_{\text{virtual}} ||\langle \psi_{\text{final}}| M_i |  \psi_{\text{virtual}} \rangle \langle \psi_{\text{virtual}} | M_j | \psi_{\text{initial}} \rangle||$ to $||\langle \psi_{\text{final}}| M_i M_j | \psi_{\text{initial}} \rangle||$  for any dipolar transition  $M_i$, $M_j$). Therefore, a mode that is active in the ($\sigma^+$, $\sigma^-$) configuration must have dominant matrix-element contributions from $||\langle \psi_{\text{final}}| J_+ J_+ | \psi_{\text{initial}} \rangle||$. So the transition that accounts for the CEF1 mode, which is active in ($\sigma^+$, $\sigma^-$), must have dominant matrix elements linked by $\langle 1| J_+ J_+ | 0 \rangle$ for a ground state wavefunction $ | 0 \rangle$ and first excited state wavefunction  $| 1 \rangle$. 

Since all the CEF modes are enhanced in the cross-circular channel and suppressed in the co-circular Raman spectra, $\langle 1| J_+ J_+ | 0 \rangle$, $\langle 1| J_- J_- | 0 \rangle$, $\langle 2| J_+ J_+ | 0 \rangle$, $\langle 2| J_- J_- | 0 \rangle$, $\langle 3| J_+ J_+ | 0 \rangle$, $\langle 3| J_- J_- | 0 \rangle$ should be large while $\langle 1| J_+ J_- | 0 \rangle$, $\langle 1| J_- J_+ | 0 \rangle$, $\langle 2| J_+ J_- | 0 \rangle$, $\langle 2| J_- J_+ | 0 \rangle$, $\langle 3| J_+ J_- | 0 \rangle$, $\langle 3| J_- J_+ | 0 \rangle$ should all be small or zero. Without the constraints imposed by these selection rules, CEF parameters consistent with the experimentally observed CEF Raman modes can be obtained regardless of the assignment of $| \psi_0 \rangle$, $| \psi_1 \rangle$, $| \psi_2 \rangle$, and $| \psi_3 \rangle$ to $| 0 \rangle$, $| 1 \rangle$, $| 2 \rangle$, and $| 3 \rangle$.  However, no parameters satisfy both the energy levels and the helicity selection rules unless $\psi^{\pm}_0$ is the ground state  $| 0 \rangle$ and the remaining modes are defined as excited states, as shown in  Figure \ref{fig:temp}b (see the Supplementary Information for further discussion). In contrast with the CEF modes, the $\omega$ mode requires $J_+ J_-$ and $J_- J_+$. Because $\omega$  comes from the parent state CEF1 and E$_{\text{2g}}$, we consider the 2$\times$2 dimensional space $ | \psi_1^\pm \rangle \otimes | \text{E}_{\text{2g}} \rangle$. Because of the strong extinction ratio between the co-circular and cross-circular channel of $\omega$, it may seem natural to conclude that $\Delta \textbf{J}_z  = 0$ for the system before and after the interaction with the photon. But in fact, the angular momentum change $\Delta \textbf{J}_z$ between the ground state and CEF1 equals $\pm 2 \hbar$ , as shown in Figure \ref{fig:eigen_sele}d, so there has to be additional angular momentum transfer due to the phonon degree of freedom. Therefore, the chiral basis \{$\text{E}_{\text{2g,+}}$, $\text{E}_{\text{2g,-}}$\} is the natural choice for describing the eigenvibration. For this 2 $\times$ 2 space, however, only 2 states, $\omega^+ = $ $| \psi_1^{-} \rangle \otimes | \text{E}_{\text{2g,+}} \rangle$ and 
$\omega^- = $ $| \psi_1^{+} \rangle \otimes | \text{E}_{\text{2g,-}} \rangle$, can be connected from the ground state by $J_+ J_-$ or $J_- J_+$. For $ | \psi_1^\pm \rangle$, the extracted $c_{1/2}$ is much smaller than $c_{7/2}$ and $c_{-{5}/{2}}$ (Methods), so we attribute the observed peaks to $|\psi^+_0 \rangle \to |\omega^- \rangle$ or $|\psi^-_0 \rangle \to |\omega^+ \rangle$.

\begin{figure*}[t]
\centering
    \includegraphics[width=1.95\columnwidth]{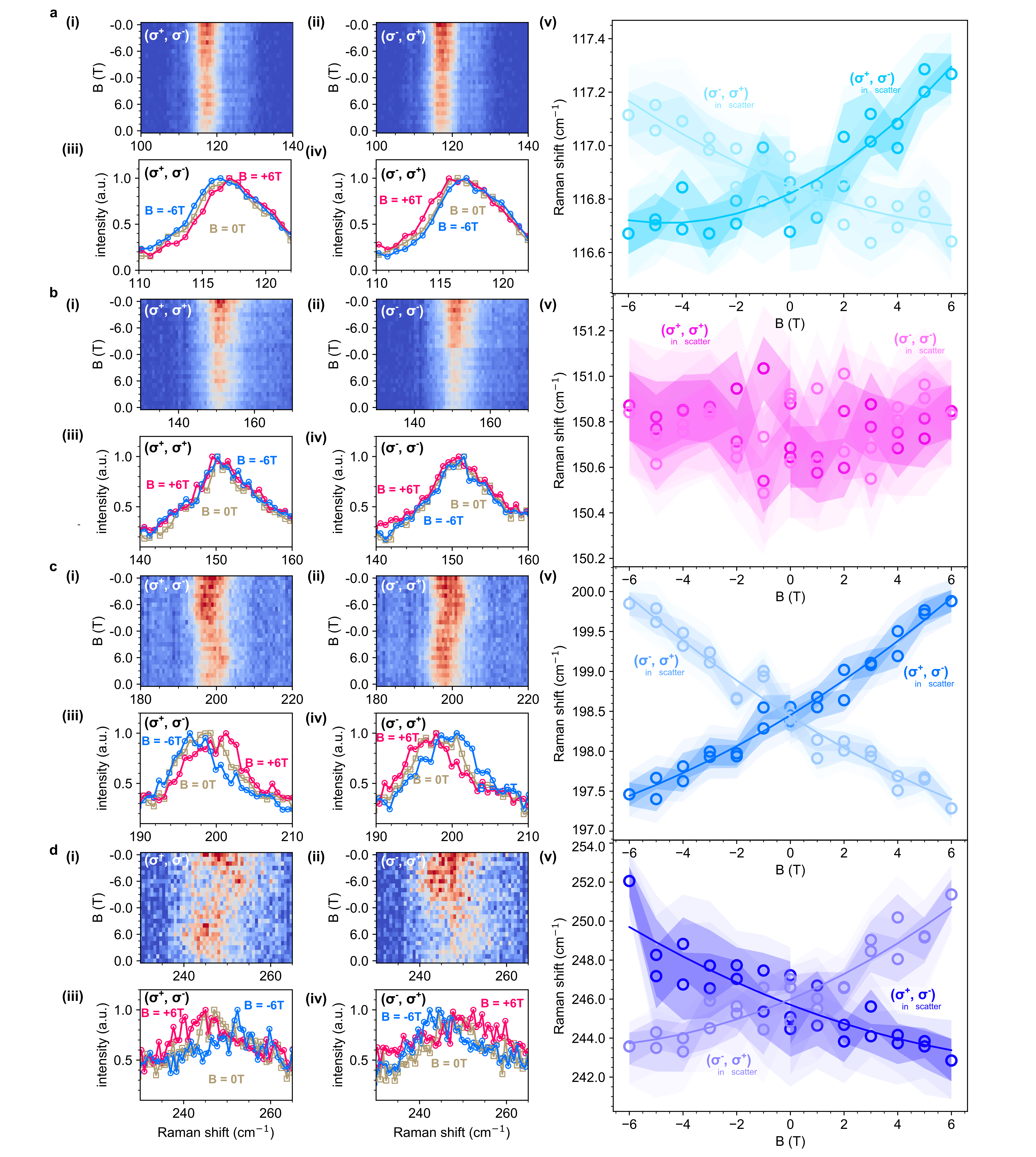}
    \caption{Helicity-resolved magnetic field dependence of CEFs and $\omega$ for NaYbSe$_2$ at $T = 4$ K. (a) CEF1. (b) $\omega$. (c) CEF2. (d) CEF3. The subplots (i), (ii) are contour plots for helicity-resolved magneto Raman with polarization configurations that the given mode is active in. (iii) and (iv) are linecuts at $B$ = 0 T, $B$ = -6 T and $B$ = +6 T. (v) peak positions extracted from (i) and (ii). The 68\% and 95\% HDIs are represented by darker and lighter shading, respectively.}
    \label{fig:field}
\end{figure*}

To distinguish the $+$ and $-$ branches, we perform magnetic field-dependent Raman spectroscopy, which lifts the degeneracy within each of the Kramers pairs and therefore provides further constraints on the CEF assignments. Magnetization and specific heat measurements have been previously used to show that no field-induced ordering is present in NaYbSe$_2$ for $\textbf{B}\; ||\; \textbf{c}$ for  $B <9$~T at $T = 4$~K~\cite{NaYbSe2_2}. Therefore, with the 6 T magnetic fields accessible in this setup, no field-induced spin ordering is expected. Figure \ref{fig:field} shows the magnetic field dependence of the CEF levels for (a) CEF1, (b) $\omega$, (c) CEF2, and (d) CEF3 measurements. Sub-panels (i, ii) are contour plots and (iii, iv) are linecuts at $B = 0~\text{T}, \;-6~\text{T},$ and $6~\text{T}$. Again, the CEF levels only show up in cross-circular polarization configurations, ($\sigma^{+}, \sigma^{-}$) and ($\sigma^{-}, \sigma^{+}$), while the $\omega$ mode is stronger in co-circular polarization configurations, ($\sigma^{+}, \sigma^{+}$) and ($\sigma^{-}, \sigma^{-}$). In the ($\sigma^{+}, \sigma^{-}$) configuration and with increasing positive magnetic field, CEF1 and CEF2 increase in energy while CEF3 decreases. This dependence is inverted if the magnetic field is reversed ($B \to -B$) or the helicity is reversed (($\sigma^{+}, \sigma^{-}$)  $\to$ ($\sigma^{-}, \sigma^{+}$)).

\begin{figure*}
\centering
    \includegraphics[width=2\columnwidth]{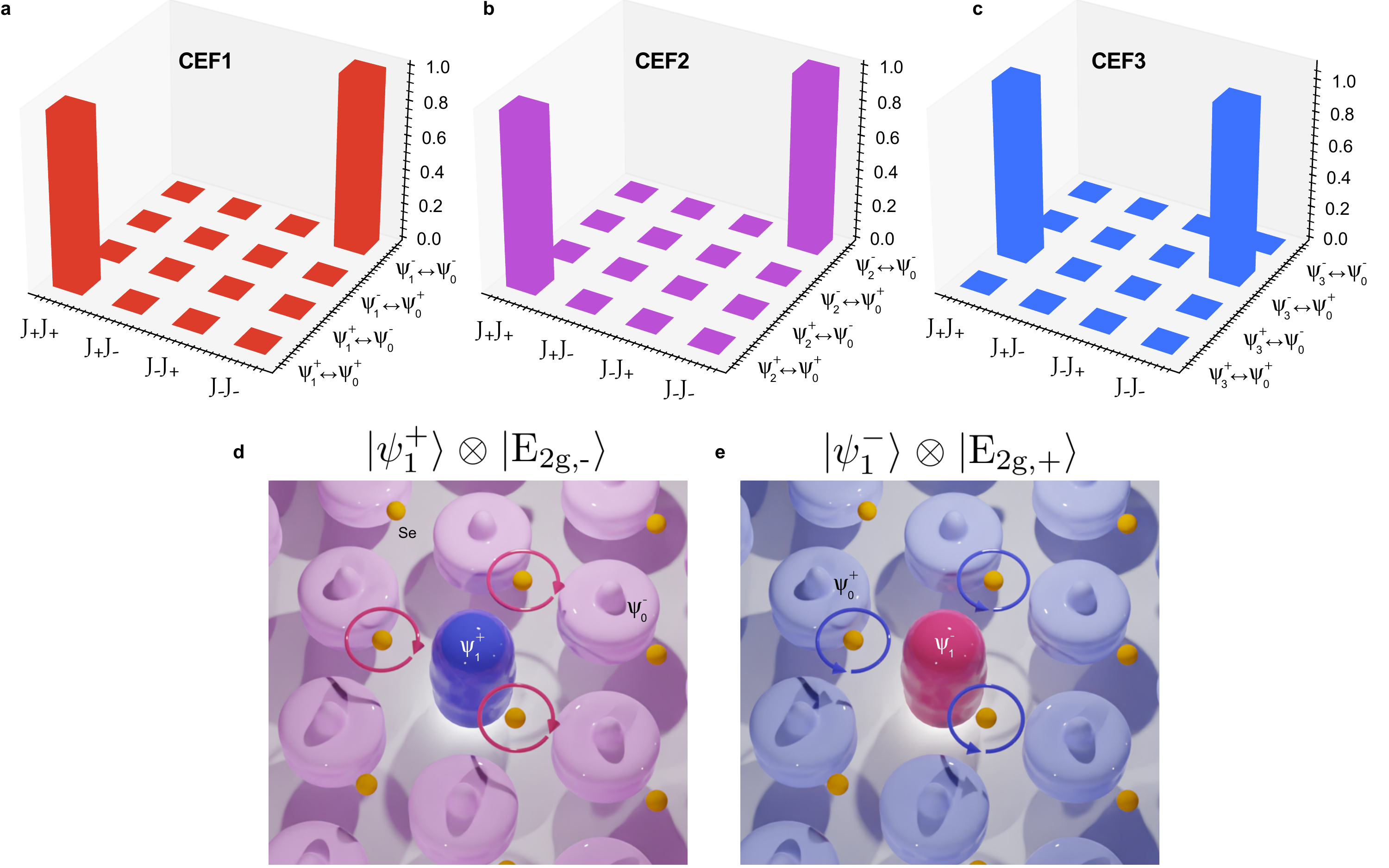}
    \caption{From helicity-resolved field dependence, transitions between the $+$ and $-$ Kramers pairs in the eigenstates $| \psi^{\pm}_0 \rangle$ and $| \psi^{\pm}_{1, 2, 3} \rangle$ can be assigned (a) CEF1 (b) CEF2 and (c) CEF3. (d) and (e) $| \psi_1^{-} \rangle \otimes | \text{E}_{\text{2g,+}} \rangle$, $| \psi_1^{+} \rangle \otimes | \text{E}_{\text{2g,-}} \rangle$.} 
    \label{fig:tomo}
\end{figure*}

Sub-panel (v) of Figure \ref{fig:field}a--d shows the Bayesian-inferred peak positions for CEF1 (a), $\omega$ (b), CEF2 (c), and CEF3 (d)  from the co- and cross-polarized spectra shown in sub-panels (i) and (ii). The aforementioned trends for the CEF modes are well-resolved in (v) while $\omega$ shows no field dependence. 

These constraints allow us to make the following assignments: for cross-circular polarized ($\sigma^{+}, \sigma^{-}$), the observed CEF1 comes from the transition $|\psi^+_0 \rangle \to |\psi^+_1 \rangle$, CEF2 comes from $|\psi^+_0 \rangle \to |\psi^+_2 \rangle$, and CEF3 comes from $|\psi^-_0 \rangle \to |\psi^+_3 \rangle$. Similarily, for cross-circular polarized ($\sigma^{-}, \sigma{+}$), the observed CEF1 comes from the transition $|\psi^-_0 \rangle \to |\psi^-_1 \rangle$, CEF2 comes from $|\psi^-_0 \rangle \to |\psi^-_2 \rangle$, and CEF3 comes from $|\psi^+_0 \rangle \to |\psi^-_3 \rangle$.  These assignments not only satisfy the aforementioned selection rule but also capture the correct sign of the magnetic field dependence when an additional Zeeman dependence $H_{\textbf{B}} = - g_J \mu_B \textbf{J}_z\cdot \textbf{B}$ is added. Figure \ref{fig:tomo}a--c shows the relative transition probability using these state assignments. The transition probabilities are consistent with the previously established set of selection rules and states (e.g., CEF1 is dominated by $|\psi^+_0 \rangle \to |\psi^+_1 \rangle$ with $J_+J_+$ and $|\psi^-_0 \rangle \to |\psi^-_1 \rangle$ with $J_-J_-$).    

Because the $\omega^\pm$ show negligible magnetic field dependence, we assign the observed transitions to $| \psi_1^{+} \rangle \to | \psi_1^{-} \rangle \otimes | \text{E}_{\text{2g,+}} \rangle$ (as shown in Figure \ref{fig:eigen_sele}d) and $| \psi_1^{-} \rangle \to | \psi_1^{+} \rangle \otimes | \text{E}_{\text{2g,-}} \rangle$. Because the transition from $| \psi_1^{+} \rangle$ to the bare $| \psi_1^{-} \rangle$ requires $\Delta J_z = +1 ~\hbar$ but the final change is $\Delta J_z = 0$, the phonon subsystem takes $ +1 \hbar$ and creates one chiral phonon $| \text{E}_{\text{2g,+}} \rangle$. Similarly, a chiral phonon $| \text{E}_{\text{2g,-}} \rangle$ is created in the transition $| \psi_1^{-} \rangle \to | \psi_1^{+} \rangle \otimes | \text{E}_{\text{2g,-}} \rangle$. Because the parent space $ | \psi_1^\pm \rangle \otimes | \text{E}_{\text{2g}} \rangle$ is 2$\times$2 dimensional, two more states $\overline{\omega}^{+} =  | \psi_1^{+} \rangle \otimes | \text{E}_{\text{2g,+}} \rangle$ and $\overline{\omega}^{-} =  | \psi_1^{-} \rangle \otimes | \text{E}_{\text{2g,-}} \rangle$ are expected, just as in the original Thalmeier-Fulde hybridization model for a VBS\cite{Thalmeier_1982}. These two states may be absent in the measured spectra because they require non-negligible $c_{1/2}$ for $| \psi_1^\pm \rangle$ to have a meaningful matrix element, as shown in the right panel of Figure \ref{fig:eigen_sele}d. This observation is surprising because NaYbSe$_2$ is $\mathcal{P}$-symmetric and $\mathcal{T}$-symmetric and because the coupling of two $\mathcal{T}$-symmetric degenerate subsystems (CEF Kramers pair $| \psi_1^{\pm} \rangle$ and $| \text{E}_{\text{2g}} \rangle$) lifts the degeneracy and block-diagonalizes the joint system into 2D subspace $\omega^\pm$ and 2D subspace $\overline{\omega}^{\pm}$. 

Figures \ref{fig:tomo}d and \ref{fig:tomo}e illustrate the 2 states, $| \psi_1^{-} \rangle \otimes | \text{E}_{\text{2g,+}} \rangle$ and $| \psi_1^{+} \rangle \otimes | \text{E}_{\text{2g,-}} \rangle$ based on the observations described above. In Figure \ref{fig:tomo}d, when the ground state is $| \psi_0^{+} \rangle$ (shown in light pink), the  $J_+ J_-$ or $J_- J_+$ operator excites the system to the $+$ branch, $| \psi_1^{+} \rangle \otimes | \text{E}_{\text{2g,-}} \rangle$ (where the CEF excitation corresponds to the $+$ branch shown in blue and the phononic excitation corresponds to $-$  angular momentum shown in dark pink). Figure \ref{fig:tomo}e shows the same effect for $| \psi_0^{+} \rangle$ to  $| \psi_1^{-} \rangle \otimes | \text{E}_{\text{2g,+}} \rangle$.  Because the transition is degenerate for $ |\psi^{\mp}_0 \rangle \to \omega^\pm$ with $J_+ J_-$ or $J_- J_+$, this eigenspace, in principle, can be transduced from a QSL ground state. If the QSL ground state is written as $\sum_j \prod_i \psi_0^{\sigma_{ij}}$ -- where $i$ is the site index, $j$ the configuration index, and $\sigma_{ij} = +$, $-$ is the branch -- then $J_+ J_-$ or $J_- J_+$ should bring the system into $\sum_j \prod_i \omega^{\sigma_{ij}}$ within the point spread volume of the excitation beam, transducing the possible ground state entanglement to the VBS $\omega^{\pm}$. 


\section{Conclusion}
We have shown here that the VBS can serve as a flexible probe of phonon chirality by measuring an angular momentum change of $\delta J_z = \pm \hbar$ in a VBS due to coupling to degenerate $E_{\text{2g}}$ phonons and modeling the interactions with a set of helicity-resolved selection rules between CEF transitions. It is natural to ask whether this coupling provides alternative pathways for spin relaxation. If the phonon band structure is modified by this coupling, it may also be possible to access the resulting Berry curvature and probe the robustness of potential topological invariants.  Additionally, the observation of photon- and CEF-mediated angular momentum transfer in a material with $\mathcal{P}$ and $\mathcal{T}$ symmetry suggests that there may be many more possible routes to the creation of chiral phonons by coupling to electronic, spin, orbital degrees of freedom in a given system.  This finding thus creates a framework that may ultimately enable the transduction of possible QSL ground states to experimentally accessible VBSs.

\section{Methods}
{\scriptsize

\subsection{Sample Details}
Single crystals of NaYbSe$_2$ were synthesized from NaCl (crystalline powder, $99+\%$), Yb (ingot, $99.9\%$), and Se (powder, $99.999\%$) via the flux method. The finely ground flux mixtures of NaCl, Yb, and Se with a molar ratio of 20:1:2.4 were heated to 850 $^{\circ}$C slowly in a vacuum. After two weeks, the furnace was cooled to room temperature at a rate of 40$^{\circ}$C/hr. The laboratory-grown single crystals of NaYbSe$_2$ were separated from excess alkali halide flux through washing with deionized water and isopropyl alcohol inside a fume-hood. CsYbSe$_2$ single crystals were grown using a related flux method that is described more detail in previous work\cite{RN3955}. 

\subsection{Raman Spectroscopy}
Variable temperature Raman spectra were acquired in a Montana Instruments closed-cycle cryostat using an in-vacuum objective with numerical aperture of 0.85. A 1.5 mW, 532.03 nm continuous wave laser excited the sample in an out-of-plane back scattering geometry (beam path $\|$ \textbf{c}). Rayleigh scattering was minimized with either a set of 3 Optigrate volume Bragg gratings or a set of Semrock RazorEdge ultrasteep dichroic and long-pass edge filters with cutoff at 90 cm$^{-1}$, and spectra were acquired with a 30 s exposure time. 

Magnetic-field-dependent Raman spectra were acquired at a fixed temperature of $T = $ 4 K for \textbf{H} $\|$ \textbf{c} in a customized Leiden dilution refrigerator with free space optical access to the sample at the mixing chamber stage\cite{lawrie2021freespace}. The spectra were taken with an Andor Kymera 193 spectrograph (2400 line/mm grating) and a Newton EMCCD DU970P-BV camera. The same laser and filters were used as with the variable temperature measurements, with the laser power set to 1.0 mW and a typical exposure time of 300 s per spectrum. Achromatic half-wave plates and quarter-wave plates were mounted on rotators for automated polarization control in both microscopes.   

\subsection{Crystal Field Hamiltonian}

For review of the approaches used here, see, for example, Bartolomé et al. \cite{BARTOLOME20171}. The CEF Hamiltonian for NaYbSe$_2$ within the point charge approximation \cite{jensen1991rare,NaYbSe2PRB_CEF,Schmidt_2021,pocs2021systematic} that describes the ground state manifold $J = 7/2$ is: 

\begin{equation}
\begin{split}
    H_{\text{CEF}} = &B_2^0 \textbf{O}_2^0 + B_4^0 \textbf{O}_4^0 + B_4^3 \textbf{O}_4^3 \\
    + &B_6^0 \textbf{O}_6^0 + B_6^3 \textbf{O}_6^3 + B_6^6 \textbf{O}_6^6\
\end{split}
\end{equation}

The expected helicity dependence of the CEF excitations follows from the selection rule bridging the two relevant states. The eigenstates for the energy levels described in equation 1 are: 
\begin{equation}
\begin{split}
    | \psi^+_{\text{1, 2, 3}} \rangle = &-\alpha e^{i \phi_\alpha}  | \frac{7}{2}, \frac{7}{2} \rangle  + \beta | \frac{7}{2}, \frac{1}{2} \rangle + \gamma e^{-i \phi_\gamma}  | \frac{7}{2}, -\frac{5}{2} \rangle\\
    | \psi^-_{\text{1, 2, 3}} \rangle = &\alpha e^{i \phi_\alpha}  | \frac{7}{2}, -\frac{7}{2} \rangle  + \beta | \frac{7}{2}, -\frac{1}{2} \rangle - \gamma e^{i \phi_\gamma}  | \frac{7}{2}, \frac{5}{2} \rangle\\
\end{split}
\end{equation}
and 
\begin{equation}
\begin{split}
    | \psi^+_0 \rangle = & | \frac{7}{2}, \frac{3}{2} \rangle  \\
    | \psi^-_0 \rangle = & | \frac{7}{2}, -\frac{3}{2} \rangle \\
\end{split}
\end{equation}
where $\alpha$, $\beta$, $\gamma$, $\phi_\alpha$, $\phi_\beta$, and $\phi_\gamma$ $\in \mathbb{R}$ are determined by the CEF parameters $B_2^0$, $B_4^0$, $B_4^3$, $B_6^0$, $B_6^3$ and $B_6^6$. The eigenstates are shown in Figure \ref{fig:eigen_sele}a. Due to the 3-fold symmetry of the Yb$^{3+}$ environment, a 3-fold periodicity and hence angular momentum folding, analogous to the the Umklapp process for linear momentum\cite{helicity_Umklapp} is expected. Without further constraints, the 6 parameters have enough degrees of freedom to fit experimentally observed energy levels. More than one set of CEF parameters that minimize the error may exist and the order of the eigenstates may not be the same across those sets. For example, the little group spanned by the special pair with single angular momentum eigenstate $ | \psi^{\pm}_0 \rangle$ is assigned to CEF1 in Zhang et al. \cite{NaYbSe2PRB_CEF} and CEF2 in Scheie et al. \cite{scheie2021witnessing}. Schimidt et al. pointed out that they cannot be the ground state due to observed in-plane field dependence at low temperatures\cite{Schmidt_2021}. In addition to the Zeeman dependence $H_{\textbf{B}} = - g_J \mu_B \textbf{J}_z\cdot \textbf{B}$, several additional corrections have been considered. For example, Pocs et al. \cite{pocs2021systematic} considered an $H_{\text{XXZ}}$ term. Zhang et al. considered anisotropic spin-spin interactions \cite{PhysRevB.103.184419.Zhang.spin.H}. 

In the context of magnetostrictive coupling with CEF modes, Callan et al. considered an additional term $H_{\text{me}} = - \sum_{\Gamma_{\nu}} \zeta(\Gamma_{\nu}) \; u(\Gamma_{\nu}) \; Q(\Gamma_{\nu})$ where $\zeta$ is the coupling strength, $u(\Gamma_{\nu})$ are phonon operators, and $Q(\Gamma_{\nu})$ is the transformed phonon mode octupolar operator on the CEF manifold. The Callen-Callen\cite{Callen, PhysRevB.61.9130} magnetoelastic interaction is quadruplar. The quadruple operator (\textit{l} = 2) is given by A$_{\text{1g}}$ + 2E$_{\text{g}}$. Hence, 
\begin{equation}
\begin{split}
H_{\text{me}} = &- \sum_{\Gamma_{\nu}} \zeta(\Gamma_{\nu}) \; u(\Gamma_{\nu}) \; Q(\Gamma_{\nu})\\
Q(A_{\text{1g}}) =& 3J_z^2 - J^2\\
Q(E_{\text{g}}^{\text{I}}) =& \alpha_1 (J_z J_x + J_x J_z) +\alpha_2 (J_y J_z + J_z J_y) \\
Q(E_{\text{g}}^{\text{II}}) =& \alpha_3 (J_x^2 - J_y^2) +\alpha_4 (J_x J_y + J_y J_x)\\
\end{split}
\end{equation}

{\setlength{\parindent}{0cm} For $Q(E_{\text{g}}^{\text{II}})$, since $J_x^2 - J_y^2 = (J_+J_+ + J_-J_-)/2$, it induces $\Delta \textbf{J}_z = \pm 2 \hbar$, while for $J_x J_y + J_y J_x = iJ_z$, $\Delta \textbf{J}_z = 0$. For $Q(E_{\text{g}}^{\text{I}})$, $J_z J_x + J_x J_z$ and $J_z J_x + J_x J_z$ both induces $\Delta \textbf{J}_z = \pm 1 \hbar$. }

\subsection{Fitting Experimental Data to CEF Hamiltonian}

To fit the experimental data, we used a general non-linear model normalization procedure in Mathematica. To start, a general set of eigenvalues and eigenvectors as a function of the CEF parameters is obtained by direct diagonalization of the CEF Hamiltonian. The eigenvectors are sorted based on their eigenvalues for each test parameter space. Then the cost function is defined by the sum of the squared errors between transitions and the data, with both inter-branch and intra-branch transitions considered. The selection rules are added as a term in the cost function when necessary. The final CEF parameters are the set that minimized the cost function.

\subsection{Bayesian Inference}

 We employed Bayesian inference to extract spectral parameters such as peak positions and widths. This approach is based on Bayes' rule: $P(\theta|y) = \frac{P(y|\theta) P(\theta) }{P(y)}$. In the context of spectral data, the priors $P(\theta)$ are the \textit{true} distribution of the parameters such as peak position, peak height, width, etc. The likelihood $P(y|\theta)$ is the experimental data. The posterior $P(\theta|y)$ is the conditional distribution of the experiment parameters given the experimental data. $P(y)$ is a normalization factor. The distribution of the priors of the parameters are assumed to be Gaussian or uniform. Additional noise, offset and slope are added to capture background unrelated to the peak parameters. To carry out the inference, the Hamiltonian Monte Carlo python package \texttt{PyMC3}\cite{pyMC} was used. A hierarchical model was constructed to concurrently extract peak parameters from a family of spectra such as the temperature dependence or magnetic field dependence.  A no U-Turns (NUTS) sampler was used with 4 chains with 3,000 samples per chain. It takes between 20 min to 3 hours for a peak over a dataset to converge.

}

\section{Data availability}
The data that support the findings of this study are available from the corresponding authors upon request. 

\acknowledgments
The authors would like to acknowledge insightful discussion with Allen Scheie, Xinshu Zhang, Yi Luo, Cristian Batista, Alan Tennant and Vyacheslav Bryantsev. This research was sponsored by the U. S. Department of Energy, Office of Science, Basic Energy Sciences, Materials Sciences and Engineering Division. The first-principles phonon calculations and variable-temperature Raman microscopy were performed at the Center for Nanophase Materials Sciences, which is a U.S. Department of Energy Office of Science User Facility. SDW and GP acknowledge support by the US Department of Energy, Office of Basic Energy Sciences, Division of Materials Sciences and Engineering under award DE-SC0017752. Postdoctoral research support was provided by the Intelligence Community Postdoctoral Research Fellowship Program at the Oak Ridge National Laboratory, administered by Oak Ridge Institute for Science and Education through an interagency agreement between the U.S. Department of Energy and the Office of the Director of National Intelligence. 

\section{Author contributions}
All authors discussed the results thoroughly. Y.-Y. P., C. E. M., and B. J. L. performed magneto-Raman measurements. L. L. performed Raman tensor analysis. G. P. and J. X. grew the samples. Y.-Y. P, L. L. did the data analysis with inputs from L. L., M. C., B. J. L.. X. L. and L. L. performed DFT calculation. A. S. S., D. P., S. W. and B. J. L initiated and oversaw the project. Y.-Y. P., L. L., B. J. L wrote most of the manuscript with contributions from all authors. 

\section{Competing interests}
\setlength{\parindent}{0cm} The authors declare no competing interests.

\section{Additional Information}
\textbf{Supplementary information} is available for this paper online. 
\textbf{Correspondence and requests for materials} should be addressed to Y.-Y. P. or B. J. L. 

\bibliographystyle{naturemag}
\bibliography{references}

\begin{thebibliography}{10}
\expandafter\ifx\csname url\endcsname\relax
  \def\url#1{\texttt{#1}}\fi
\expandafter\ifx\csname urlprefix\endcsname\relax\def\urlprefix{URL }\fi
\providecommand{\bibinfo}[2]{#2}
\providecommand{\eprint}[2][]{\url{#2}}

\bibitem{Bloch1929}
\bibinfo{author}{Bloch, F.}
\newblock \bibinfo{title}{Über die quantenmechanik der elektronen in
  kristallgittern}.
\newblock \emph{\bibinfo{journal}{Zeitschrift für Physik}}
  \textbf{\bibinfo{volume}{52}}, \bibinfo{pages}{555--600}
  (\bibinfo{year}{1929}).
\newblock \urlprefix\url{https://doi.org/10.1007/BF01339455}.

\bibitem{chiralphonon_2D:review}
\bibinfo{author}{Chen, H.}, \bibinfo{author}{Zhang, W.}, \bibinfo{author}{Niu,
  Q.} \& \bibinfo{author}{Zhang, L.}
\newblock \bibinfo{title}{Chiral phonons in two-dimensional materials}.
\newblock \emph{\bibinfo{journal}{2D Materials}} \textbf{\bibinfo{volume}{6}},
  \bibinfo{pages}{012002} (\bibinfo{year}{2018}).
\newblock \urlprefix\url{https://doi.org/10.1088/2053-1583/aaf292}.

\bibitem{PhysRevLett.115.115502}
\bibinfo{author}{Zhang, L.} \& \bibinfo{author}{Niu, Q.}
\newblock \bibinfo{title}{Chiral phonons at high-symmetry points in monolayer
  hexagonal lattices}.
\newblock \emph{\bibinfo{journal}{Phys. Rev. Lett.}}
  \textbf{\bibinfo{volume}{115}}, \bibinfo{pages}{115502}
  (\bibinfo{year}{2015}).
\newblock
  \urlprefix\url{https://link.aps.org/doi/10.1103/PhysRevLett.115.115502}.

\bibitem{JUNEJA2021100548}
\bibinfo{author}{Juneja, R.} \emph{et~al.}
\newblock \bibinfo{title}{Quasiparticle twist dynamics in non-symmorphic
  materials}.
\newblock \emph{\bibinfo{journal}{Materials Today Physics}}
  \textbf{\bibinfo{volume}{21}}, \bibinfo{pages}{100548}
  (\bibinfo{year}{2021}).
\newblock
  \urlprefix\url{https://www.sciencedirect.com/science/article/pii/S2542529321002091}.

\bibitem{PhysRevB.98.241405}
\bibinfo{author}{Pandey, T.}, \bibinfo{author}{Polanco, C.~A.},
  \bibinfo{author}{Cooper, V.~R.}, \bibinfo{author}{Parker, D.~S.} \&
  \bibinfo{author}{Lindsay, L.}
\newblock \bibinfo{title}{Symmetry-driven phonon chirality and transport in
  one-dimensional and bulk $\mathrm{B}{\mathrm{a}}_{3}\mathrm{N}$-derived
  materials}.
\newblock \emph{\bibinfo{journal}{Phys. Rev. B}} \textbf{\bibinfo{volume}{98}},
  \bibinfo{pages}{241405} (\bibinfo{year}{2018}).
\newblock \urlprefix\url{https://link.aps.org/doi/10.1103/PhysRevB.98.241405}.

\bibitem{PhysRevMaterials.5.085002}
\bibinfo{author}{Gunatilleke, W. D. C.~B.} \emph{et~al.}
\newblock \bibinfo{title}{Intrinsic anharmonicity and thermal properties of
  ultralow thermal conductivity
  ${\mathrm{ba}}_{6}{\mathrm{sn}}_{6}{\mathrm{se}}_{13}$}.
\newblock \emph{\bibinfo{journal}{Phys. Rev. Materials}}
  \textbf{\bibinfo{volume}{5}}, \bibinfo{pages}{085002} (\bibinfo{year}{2021}).
\newblock
  \urlprefix\url{https://link.aps.org/doi/10.1103/PhysRevMaterials.5.085002}.

\bibitem{coh2021classification}
\bibinfo{author}{Coh, S.}
\newblock \bibinfo{title}{Classification of materials with phonon angular
  momentum and microscopic origin of angular momentum} (\bibinfo{year}{2021}).
\newblock \eprint{1911.05064}.

\bibitem{chiralphonon}
\bibinfo{author}{Zhu, H.} \emph{et~al.}
\newblock \bibinfo{title}{Observation of chiral phonons}.
\newblock \emph{\bibinfo{journal}{Science}} \textbf{\bibinfo{volume}{359}},
  \bibinfo{pages}{579--582} (\bibinfo{year}{2018}).
\newblock
  \urlprefix\url{https://www.science.org/doi/abs/10.1126/science.aar2711}.
\newblock \eprint{https://www.science.org/doi/pdf/10.1126/science.aar2711}.

\bibitem{Chiral_CrBr3}
\bibinfo{author}{Yin, T.} \emph{et~al.}
\newblock \bibinfo{title}{Chiral phonons and giant magneto-optical effect in
  crbr3 2d magnet}.
\newblock \emph{\bibinfo{journal}{Advanced Materials}}
  \textbf{\bibinfo{volume}{33}}, \bibinfo{pages}{2101618}
  (\bibinfo{year}{2021}).

\bibitem{manenkov1966spin}
\bibinfo{author}{Manenkov, A.~A.} \& \bibinfo{author}{Orbach, R.}
\newblock \emph{\bibinfo{title}{Spin-lattice relaxation in ionic solids}}
  (\bibinfo{publisher}{Harper \& Row}, \bibinfo{year}{1966}).

\bibitem{deHaas2}
\bibinfo{author}{Nakane, J.~J.} \& \bibinfo{author}{Kohno, H.}
\newblock \bibinfo{title}{Angular momentum of phonons and its application to
  single-spin relaxation}.
\newblock \emph{\bibinfo{journal}{Phys. Rev. B}} \textbf{\bibinfo{volume}{97}},
  \bibinfo{pages}{174403} (\bibinfo{year}{2018}).
\newblock \urlprefix\url{https://link.aps.org/doi/10.1103/PhysRevB.97.174403}.

\bibitem{PAM}
\bibinfo{author}{Streib, S.}
\newblock \bibinfo{title}{Difference between angular momentum and pseudoangular
  momentum}.
\newblock \emph{\bibinfo{journal}{Phys. Rev. B}}
  \textbf{\bibinfo{volume}{103}}, \bibinfo{pages}{L100409}
  (\bibinfo{year}{2021}).
\newblock
  \urlprefix\url{https://link.aps.org/doi/10.1103/PhysRevB.103.L100409}.

\bibitem{PhononHall}
\bibinfo{author}{Strohm, C.}, \bibinfo{author}{Rikken, G. L. J.~A.} \&
  \bibinfo{author}{Wyder, P.}
\newblock \bibinfo{title}{Phenomenological evidence for the phonon hall
  effect}.
\newblock \emph{\bibinfo{journal}{Phys. Rev. Lett.}}
  \textbf{\bibinfo{volume}{95}}, \bibinfo{pages}{155901}
  (\bibinfo{year}{2005}).
\newblock
  \urlprefix\url{https://link.aps.org/doi/10.1103/PhysRevLett.95.155901}.

\bibitem{EinsteinDeHaas}
\bibinfo{author}{Zhang, L.} \& \bibinfo{author}{Niu, Q.}
\newblock \bibinfo{title}{Angular momentum of phonons and the einstein--de haas
  effect}.
\newblock \emph{\bibinfo{journal}{Phys. Rev. Lett.}}
  \textbf{\bibinfo{volume}{112}}, \bibinfo{pages}{085503}
  (\bibinfo{year}{2014}).
\newblock
  \urlprefix\url{https://link.aps.org/doi/10.1103/PhysRevLett.112.085503}.

\bibitem{circularHeat}
\bibinfo{author}{Qin, T.}, \bibinfo{author}{Niu, Q.} \& \bibinfo{author}{Shi,
  J.}
\newblock \bibinfo{title}{Energy magnetization and the thermal hall effect}.
\newblock \emph{\bibinfo{journal}{Phys. Rev. Lett.}}
  \textbf{\bibinfo{volume}{107}}, \bibinfo{pages}{236601}
  (\bibinfo{year}{2011}).
\newblock
  \urlprefix\url{https://link.aps.org/doi/10.1103/PhysRevLett.107.236601}.

\bibitem{Chen19258}
\bibinfo{author}{Chen, Z.}, \bibinfo{author}{Luo, J.-W.} \&
  \bibinfo{author}{Wang, L.-W.}
\newblock \bibinfo{title}{Revealing angular momentum transfer channels and
  timescales in the ultrafast demagnetization process of ferromagnetic
  semiconductors}.
\newblock \emph{\bibinfo{journal}{Proceedings of the National Academy of
  Sciences}} \textbf{\bibinfo{volume}{116}}, \bibinfo{pages}{19258--19263}
  (\bibinfo{year}{2019}).
\newblock \urlprefix\url{https://www.pnas.org/content/116/39/19258}.
\newblock \eprint{https://www.pnas.org/content/116/39/19258.full.pdf}.

\bibitem{ultraDemag}
\bibinfo{author}{Tsatsoulis, T.} \emph{et~al.}
\newblock \bibinfo{title}{Ultrafast demagnetization after femtosecond laser
  pulses: Transfer of angular momentum from the electronic system to
  magnetoelastic spin-phonon modes}.
\newblock \emph{\bibinfo{journal}{Phys. Rev. B}} \textbf{\bibinfo{volume}{93}},
  \bibinfo{pages}{134411} (\bibinfo{year}{2016}).
\newblock \urlprefix\url{https://link.aps.org/doi/10.1103/PhysRevB.93.134411}.

\bibitem{MottSc_NaYbSe2}
\bibinfo{author}{Jia, Y.-T.} \emph{et~al.}
\newblock \bibinfo{title}{Mott transition and superconductivity in quantum spin
  liquid candidate ${{NaYbSe}}_{2}$}.
\newblock \emph{\bibinfo{journal}{Chinese Physics Letters}}
  \textbf{\bibinfo{volume}{37}}, \bibinfo{pages}{097404}
  (\bibinfo{year}{2020}).
\newblock \urlprefix\url{http://dx.doi.org/10.1088/0256-307X/37/9/097404}.

\bibitem{NaYbSe2_PRX}
\bibinfo{author}{Dai, P.-L.} \emph{et~al.}
\newblock \bibinfo{title}{Spinon fermi surface spin liquid in a triangular
  lattice antiferromagnet ${{NaYbSe}}_{2}$}.
\newblock \emph{\bibinfo{journal}{Phys. Rev. X}} \textbf{\bibinfo{volume}{11}},
  \bibinfo{pages}{021044} (\bibinfo{year}{2021}).
\newblock \urlprefix\url{https://link.aps.org/doi/10.1103/PhysRevX.11.021044}.

\bibitem{NaYbSe2_2}
\bibinfo{author}{Ranjith, K.~M.} \emph{et~al.}
\newblock \bibinfo{title}{Anisotropic field-induced ordering in the
  triangular-lattice quantum spin liquid ${{NaYbSe}}_{2}$}.
\newblock \emph{\bibinfo{journal}{Phys. Rev. B}}
  \textbf{\bibinfo{volume}{100}}, \bibinfo{pages}{224417}
  (\bibinfo{year}{2019}).
\newblock \urlprefix\url{https://link.aps.org/doi/10.1103/PhysRevB.100.224417}.

\bibitem{NaYbSe2PRB_CEF}
\bibinfo{author}{Zhang, Z.} \emph{et~al.}
\newblock \bibinfo{title}{Crystalline electric field excitations in the quantum
  spin liquid candidate ${{NaYbSe}}_{2}$}.
\newblock \emph{\bibinfo{journal}{Phys. Rev. B}}
  \textbf{\bibinfo{volume}{103}}, \bibinfo{pages}{035144}
  (\bibinfo{year}{2021}).
\newblock \urlprefix\url{https://link.aps.org/doi/10.1103/PhysRevB.103.035144}.

\bibitem{zhang2020pressure}
\bibinfo{author}{Zhang, Z.} \emph{et~al.}
\newblock \bibinfo{title}{Pressure induced metallization and possible
  unconventional superconductivity in spin liquid $naybse_{2}$}
  (\bibinfo{year}{2020}).
\newblock \eprint{2003.11479}.

\bibitem{ANDERSON1973153}
\bibinfo{author}{Anderson, P.}
\newblock \bibinfo{title}{Resonating valence bonds: A new kind of insulator?}
\newblock \emph{\bibinfo{journal}{Materials Research Bulletin}}
  \textbf{\bibinfo{volume}{8}}, \bibinfo{pages}{153--160}
  (\bibinfo{year}{1973}).
\newblock
  \urlprefix\url{https://www.sciencedirect.com/science/article/pii/0025540873901670}.

\bibitem{Knolle2019}
\bibinfo{author}{Knolle, J.} \& \bibinfo{author}{Moessner, R.}
\newblock \bibinfo{title}{A field guide to spin liquids}.
\newblock \emph{\bibinfo{journal}{Annual Review of Condensed Matter Physics}}
  \textbf{\bibinfo{volume}{10}}, \bibinfo{pages}{451--472}
  (\bibinfo{year}{2019}).
\newblock
  \urlprefix\url{https://doi.org/10.1146/annurev-conmatphys-031218-013401}.

\bibitem{Savary2016}
\bibinfo{author}{Savary, L.} \& \bibinfo{author}{Balents, L.}
\newblock \bibinfo{title}{Quantum spin liquids: a review}.
\newblock \emph{\bibinfo{journal}{Reports on Progress in Physics}}
  \textbf{\bibinfo{volume}{80}}, \bibinfo{pages}{016502}
  (\bibinfo{year}{2016}).
\newblock \urlprefix\url{http://dx.doi.org/10.1088/0034-4885/80/1/016502}.

\bibitem{barkeshli2014coherent}
\bibinfo{author}{Barkeshli, M.}, \bibinfo{author}{Berg, E.} \&
  \bibinfo{author}{Kivelson, S.}
\newblock \bibinfo{title}{Coherent transmutation of electrons into
  fractionalized anyons}.
\newblock \emph{\bibinfo{journal}{Science}} \textbf{\bibinfo{volume}{346}},
  \bibinfo{pages}{722--725} (\bibinfo{year}{2014}).

\bibitem{Schmidt_2021}
\bibinfo{author}{Schmidt, B.}, \bibinfo{author}{Sichelschmidt, J.},
  \bibinfo{author}{Ranjith, K.~M.}, \bibinfo{author}{Doert, T.} \&
  \bibinfo{author}{Baenitz, M.}
\newblock \bibinfo{title}{Yb delafossites: Unique exchange frustration of $4f$
  spin-$\frac{1}{2}$ moments on a perfect triangular lattice}.
\newblock \emph{\bibinfo{journal}{Phys. Rev. B}}
  \textbf{\bibinfo{volume}{103}}, \bibinfo{pages}{214445}
  (\bibinfo{year}{2021}).
\newblock \urlprefix\url{https://link.aps.org/doi/10.1103/PhysRevB.103.214445}.

\bibitem{NaYbS2_2018}
\bibinfo{author}{Baenitz, M.} \emph{et~al.}
\newblock \bibinfo{title}{${{NaYbS}}_{2}$: A planar spin-$\frac{1}{2}$
  triangular-lattice magnet and putative spin liquid}.
\newblock \emph{\bibinfo{journal}{Phys. Rev. B}} \textbf{\bibinfo{volume}{98}},
  \bibinfo{pages}{220409} (\bibinfo{year}{2018}).
\newblock \urlprefix\url{https://link.aps.org/doi/10.1103/PhysRevB.98.220409}.

\bibitem{NaYbS2_2019}
\bibinfo{author}{Sarkar, R.} \emph{et~al.}
\newblock \bibinfo{title}{Quantum spin liquid ground state in the disorder free
  triangular lattice ${{NaYbS}}_{2}$}.
\newblock \emph{\bibinfo{journal}{Phys. Rev. B}}
  \textbf{\bibinfo{volume}{100}}, \bibinfo{pages}{241116}
  (\bibinfo{year}{2019}).
\newblock \urlprefix\url{https://link.aps.org/doi/10.1103/PhysRevB.100.241116}.

\bibitem{NaYbO2_1}
\bibinfo{author}{Ding, L.} \emph{et~al.}
\newblock \bibinfo{title}{Gapless spin-liquid state in the structurally
  disorder-free triangular antiferromagnet ${{NaYbO}}_{2}$}.
\newblock \emph{\bibinfo{journal}{Phys. Rev. B}}
  \textbf{\bibinfo{volume}{100}}, \bibinfo{pages}{144432}
  (\bibinfo{year}{2019}).
\newblock \urlprefix\url{https://link.aps.org/doi/10.1103/PhysRevB.100.144432}.

\bibitem{NaYbO2_3}
\bibinfo{author}{Ranjith, K.~M.} \emph{et~al.}
\newblock \bibinfo{title}{Field-induced instability of the quantum spin liquid
  ground state in the ${J}_{{eff}}=\frac{1}{2}$ triangular-lattice compound
  ${{NaYbO}}_{2}$}.
\newblock \emph{\bibinfo{journal}{Phys. Rev. B}} \textbf{\bibinfo{volume}{99}},
  \bibinfo{pages}{180401} (\bibinfo{year}{2019}).
\newblock \urlprefix\url{https://link.aps.org/doi/10.1103/PhysRevB.99.180401}.

\bibitem{NaYbO2_SDW}
\bibinfo{author}{Bordelon, M.~M.} \emph{et~al.}
\newblock \bibinfo{title}{Field-tunable quantum disordered ground state in the
  triangular-lattice antiferromagnet naybo2}.
\newblock \emph{\bibinfo{journal}{Nature Physics}}
  \textbf{\bibinfo{volume}{15}}, \bibinfo{pages}{1058--1064}
  (\bibinfo{year}{2019}).
\newblock \urlprefix\url{https://doi.org/10.1038/s41567-019-0594-5}.

\bibitem{Xing_Jie_PRB2019Rapid}
\bibinfo{author}{Xing, J.} \emph{et~al.}
\newblock \bibinfo{title}{Field-induced magnetic transition and spin
  fluctuations in the quantum spin-liquid candidate ${{CsYbSe}}_{2}$}.
\newblock \emph{\bibinfo{journal}{Phys. Rev. B}}
  \textbf{\bibinfo{volume}{100}}, \bibinfo{pages}{220407}
  (\bibinfo{year}{2019}).
\newblock \urlprefix\url{https://link.aps.org/doi/10.1103/PhysRevB.100.220407}.

\bibitem{pocs2021systematic}
\bibinfo{author}{Pocs, C.~A.} \emph{et~al.}
\newblock \bibinfo{title}{Systematic fitting of crystal-field levels and
  accurate extraction of quantum magnetic models in triangular-lattice
  delafossites} (\bibinfo{year}{2021}).

\bibitem{xie2021fieldinduced}
\bibinfo{author}{Xie, T.} \emph{et~al.}
\newblock \bibinfo{title}{Field-induced spin excitations in the spin-1/2
  triangular-lattice antiferromagnet csybse$_2$} (\bibinfo{year}{2021}).
\newblock \eprint{2106.12451}.

\bibitem{RN3955}
\bibinfo{author}{Xing, J.} \emph{et~al.}
\newblock \bibinfo{title}{Crystal synthesis and frustrated magnetism in
  triangular lattice ${CsRESe_2}$ (re = la–lu): Quantum spin liquid
  candidates ${CsCeSe_2}$ and ${CsYbSe_2}$}.
\newblock \emph{\bibinfo{journal}{ACS Materials Letters}}
  \textbf{\bibinfo{volume}{2}}, \bibinfo{pages}{71--75} (\bibinfo{year}{2020}).
\newblock \urlprefix\url{https://doi.org/10.1021/acsmaterialslett.9b00464}.

\bibitem{pai2021nearlyresonant}
\bibinfo{author}{Pai, Y.-Y.} \emph{et~al.}
\newblock \bibinfo{title}{Mesoscale interplay between phonons and crystal
  electric field excitations in quantum spin liquid candidate csybse2}.
\newblock \emph{\bibinfo{journal}{J. Mater. Chem. C}} \bibinfo{pages}{--}
  (\bibinfo{year}{2022}).
\newblock \urlprefix\url{http://dx.doi.org/10.1039/D1TC05934C}.

\bibitem{xing2021synthesis}
\bibinfo{author}{Xing, J.}, \bibinfo{author}{Sanjeewa, L.~D.},
  \bibinfo{author}{May, A.~F.} \& \bibinfo{author}{Sefat, A.~S.}
\newblock \bibinfo{title}{Synthesis and anisotropic magnetism in quantum spin
  liquid candidates {$A$YbSe$_2$} ({$A$ = K and Rb})} (\bibinfo{year}{2021}).
\newblock \eprint{2109.00384}.

\bibitem{scheie2021witnessing}
\bibinfo{author}{Scheie, A.~O.} \emph{et~al.}
\newblock \bibinfo{title}{Witnessing quantum criticality and entanglement in
  the triangular antiferromagnet kybse$_2$} (\bibinfo{year}{2021}).
\newblock \eprint{2109.11527}.

\bibitem{Koningstein_1967}
\bibinfo{author}{Koningstein, J.~A.}
\newblock \bibinfo{title}{Electronic raman spectra. i. raman transitions of
  trivalent ytterbium, europium, and neodymium in yttrium gallium garnet}.
\newblock \emph{\bibinfo{journal}{The Journal of Chemical Physics}}
  \textbf{\bibinfo{volume}{46}}, \bibinfo{pages}{2811--2816}
  (\bibinfo{year}{1967}).
\newblock \urlprefix\url{https://doi.org/10.1063/1.1841119}.
\newblock \eprint{https://doi.org/10.1063/1.1841119}.

\bibitem{Petit}
\bibinfo{author}{Petit, V.}, \bibinfo{author}{Camy, P.},
  \bibinfo{author}{Doualan, J.-L.}, \bibinfo{author}{Portier, X.} \&
  \bibinfo{author}{Moncorg\'e, R.}
\newblock \bibinfo{title}{Spectroscopy of ${\text{yb}}^{3+}:{\text{caf}}_{2}$:
  From isolated centers to clusters}.
\newblock \emph{\bibinfo{journal}{Phys. Rev. B}} \textbf{\bibinfo{volume}{78}},
  \bibinfo{pages}{085131} (\bibinfo{year}{2008}).
\newblock \urlprefix\url{https://link.aps.org/doi/10.1103/PhysRevB.78.085131}.

\bibitem{k6695}
\bibinfo{author}{{\v C}erm{\'a}k, P.} \emph{et~al.}
\newblock \bibinfo{title}{Magnetoelastic hybrid excitations in ${CeAuAl_3}$}.
\newblock \emph{\bibinfo{journal}{Proceedings of the National Academy of
  Sciences}} \textbf{\bibinfo{volume}{116}}, \bibinfo{pages}{6695--6700}
  (\bibinfo{year}{2019}).

\bibitem{Sethi_PRL_2019_omegas}
\bibinfo{author}{Sethi, A.}, \bibinfo{author}{Slimak, J.~E.},
  \bibinfo{author}{Kolodiazhnyi, T.} \& \bibinfo{author}{Cooper, S.~L.}
\newblock \bibinfo{title}{Emergent vibronic excitations in the
  magnetodielectric regime of ${{Ce}}_{2}{{O}}_{3}$}.
\newblock \emph{\bibinfo{journal}{Phys. Rev. Lett.}}
  \textbf{\bibinfo{volume}{122}}, \bibinfo{pages}{177601}
  (\bibinfo{year}{2019}).
\newblock
  \urlprefix\url{https://link.aps.org/doi/10.1103/PhysRevLett.122.177601}.

\bibitem{JTCeAl2}
\bibinfo{author}{Loewenhaupt, M.}, \bibinfo{author}{Rainford, B.~D.} \&
  \bibinfo{author}{Steglich, F.}
\newblock \bibinfo{title}{Dynamic jahn-teller effect in a rare-earth compound:
  Ce${{Al}}_{2}$}.
\newblock \emph{\bibinfo{journal}{Phys. Rev. Lett.}}
  \textbf{\bibinfo{volume}{42}}, \bibinfo{pages}{1709--1712}
  (\bibinfo{year}{1979}).
\newblock \urlprefix\url{https://link.aps.org/doi/10.1103/PhysRevLett.42.1709}.

\bibitem{RamanCeAl2}
\bibinfo{author}{G\"untherodt, G.}, \bibinfo{author}{Jayaraman, A.},
  \bibinfo{author}{Batlogg, G.}, \bibinfo{author}{Croft, M.} \&
  \bibinfo{author}{Melczer, E.}
\newblock \bibinfo{title}{Raman scattering from coupled phonon and electronic
  crystal-field excitations in {Ce${{Al}}_{2}$}}.
\newblock \emph{\bibinfo{journal}{Phys. Rev. Lett.}}
  \textbf{\bibinfo{volume}{51}}, \bibinfo{pages}{2330--2332}
  (\bibinfo{year}{1983}).
\newblock \urlprefix\url{https://link.aps.org/doi/10.1103/PhysRevLett.51.2330}.

\bibitem{Thalmeier_1982}
\bibinfo{author}{Thalmeier, P.} \& \bibinfo{author}{Fulde, P.}
\newblock \bibinfo{title}{Bound state between a crystal-field excitation and a
  phonon in {Ce${{Al}}_{2}$}}.
\newblock \emph{\bibinfo{journal}{Phys. Rev. Lett.}}
  \textbf{\bibinfo{volume}{49}}, \bibinfo{pages}{1588--1591}
  (\bibinfo{year}{1982}).
\newblock \urlprefix\url{https://link.aps.org/doi/10.1103/PhysRevLett.49.1588}.

\bibitem{adroja2012vibron}
\bibinfo{author}{Adroja, D.} \emph{et~al.}
\newblock \bibinfo{title}{Vibron quasibound state in the noncentrosymmetric
  tetragonal heavy-fermion compound {$CeCuAl_3$}}.
\newblock \emph{\bibinfo{journal}{Physical review letters}}
  \textbf{\bibinfo{volume}{108}}, \bibinfo{pages}{216402}
  (\bibinfo{year}{2012}).

\bibitem{CEFPhononPrNi5}
\bibinfo{author}{Aksenov, V.}, \bibinfo{author}{Goremychkin, E.},
  \bibinfo{author}{Mühle, E.}, \bibinfo{author}{Frauenheim, T.} \&
  \bibinfo{author}{Bührer, W.}
\newblock \bibinfo{title}{Coupled quadrupole-phonon excitations: Inelastic
  neutron scattering on van vleck paramagnet ${PrNi_5}$}.
\newblock \emph{\bibinfo{journal}{Physica B+C}} \textbf{\bibinfo{volume}{120}},
  \bibinfo{pages}{310--313} (\bibinfo{year}{1983}).
\newblock
  \urlprefix\url{https://www.sciencedirect.com/science/article/pii/0378436383903984}.

\bibitem{CEF_highTC}
\bibinfo{author}{Heyen, E.~T.}, \bibinfo{author}{Wegerer, R.} \&
  \bibinfo{author}{Cardona, M.}
\newblock \bibinfo{title}{Coupling of phonons to crystal-field excitations in
  ${{NdBa}}_{2}$${{Cu}}_{3}$${{O}}_{7{\ensuremath{-}}{\ensuremath{\delta}}}$}.
\newblock \emph{\bibinfo{journal}{Phys. Rev. Lett.}}
  \textbf{\bibinfo{volume}{67}}, \bibinfo{pages}{144--147}
  (\bibinfo{year}{1991}).
\newblock \urlprefix\url{https://link.aps.org/doi/10.1103/PhysRevLett.67.144}.

\bibitem{Gaudet_CEF_ph_Ho2Ti2O7}
\bibinfo{author}{Gaudet, J.} \emph{et~al.}
\newblock \bibinfo{title}{Magnetoelastically induced vibronic bound state in
  the spin-ice pyrochlore ${{Ho}}_{2}{{Ti}}_{2}{{O}}_{7}$}.
\newblock \emph{\bibinfo{journal}{Phys. Rev. B}} \textbf{\bibinfo{volume}{98}},
  \bibinfo{pages}{014419} (\bibinfo{year}{2018}).
\newblock \urlprefix\url{https://link.aps.org/doi/10.1103/PhysRevB.98.014419}.

\bibitem{CEFEatenConfigCrossOver}
\bibinfo{author}{G\"untherodt, G.}, \bibinfo{author}{Jayaraman, A.},
  \bibinfo{author}{Anastassakis, E.}, \bibinfo{author}{Bucher, E.} \&
  \bibinfo{author}{Bach, H.}
\newblock \bibinfo{title}{Effect of configuration crossover on the electronic
  {Raman} scattering by $4f$ multiplets}.
\newblock \emph{\bibinfo{journal}{Phys. Rev. Lett.}}
  \textbf{\bibinfo{volume}{46}}, \bibinfo{pages}{855--858}
  (\bibinfo{year}{1981}).
\newblock \urlprefix\url{https://link.aps.org/doi/10.1103/PhysRevLett.46.855}.

\bibitem{zhu2021fluctuating}
\bibinfo{author}{Zhu, Z.~H.} \emph{et~al.}
\newblock \bibinfo{title}{Fluctuating magnetic droplets immersed in a sea of
  quantum spin liquid} (\bibinfo{year}{2021}).
\newblock \eprint{2112.06523}.

\bibitem{helicity_Umklapp}
\bibinfo{author}{Tatsumi, Y.}, \bibinfo{author}{Kaneko, T.} \&
  \bibinfo{author}{Saito, R.}
\newblock \bibinfo{title}{Conservation law of angular momentum in
  helicity-dependent raman and rayleigh scattering}.
\newblock \emph{\bibinfo{journal}{Phys. Rev. B}} \textbf{\bibinfo{volume}{97}},
  \bibinfo{pages}{195444} (\bibinfo{year}{2018}).
\newblock \urlprefix\url{https://link.aps.org/doi/10.1103/PhysRevB.97.195444}.

\bibitem{pyMC}
\bibinfo{author}{Salvatier, J.}, \bibinfo{author}{Wiecki, T.~V.} \&
  \bibinfo{author}{Fonnesbeck, C.}
\newblock \bibinfo{title}{Probabilistic programming in python using pymc3}.
\newblock \emph{\bibinfo{journal}{PeerJ Computer Science}}
  \textbf{\bibinfo{volume}{2}}, \bibinfo{pages}{e55} (\bibinfo{year}{2016}).
\newblock \urlprefix\url{https://doi.org/10.7717/peerj-cs.55}.

\bibitem{Deng:wm6047}
\bibinfo{author}{Deng, B.} \& \bibinfo{author}{Ibers, J.~A.}
\newblock \bibinfo{title}{{CsYbSe${\sb 2}$}}.
\newblock \emph{\bibinfo{journal}{Acta Crystallographica Section E}}
  \textbf{\bibinfo{volume}{61}}, \bibinfo{pages}{i15--i17}
  (\bibinfo{year}{2005}).
\newblock \urlprefix\url{https://doi.org/10.1107/S1600536805001157}.

\bibitem{Helicity}
\bibinfo{author}{Alpeggiani, F.}, \bibinfo{author}{Bliokh, K.~Y.},
  \bibinfo{author}{Nori, F.} \& \bibinfo{author}{Kuipers, L.}
\newblock \bibinfo{title}{Electromagnetic helicity in complex media}.
\newblock \emph{\bibinfo{journal}{Phys. Rev. Lett.}}
  \textbf{\bibinfo{volume}{120}}, \bibinfo{pages}{243605}
  (\bibinfo{year}{2018}).
\newblock
  \urlprefix\url{https://link.aps.org/doi/10.1103/PhysRevLett.120.243605}.

\bibitem{Closure}
\bibinfo{author}{Axe, J.~D.}
\newblock \bibinfo{title}{Two-photon processes in complex atoms}.
\newblock \emph{\bibinfo{journal}{Phys. Rev.}} \textbf{\bibinfo{volume}{136}},
  \bibinfo{pages}{A42--A45} (\bibinfo{year}{1964}).
\newblock \urlprefix\url{https://link.aps.org/doi/10.1103/PhysRev.136.A42}.

\bibitem{lawrie2021freespace}
\bibinfo{author}{Lawrie, B.~J.}, \bibinfo{author}{Feldman, M.},
  \bibinfo{author}{Marvinney, C.~E.} \& \bibinfo{author}{Pai, Y.~Y.}
\newblock \bibinfo{title}{Free-space confocal magneto-optical spectroscopies at
  millikelvin temperatures} (\bibinfo{year}{2021}).
\newblock \eprint{2103.06851}.

\bibitem{BARTOLOME20171}
\bibinfo{author}{Bartolomé, E.}, \bibinfo{author}{Arauzo, A.},
  \bibinfo{author}{Luzón, J.}, \bibinfo{author}{Bartolomé, J.} \&
  \bibinfo{author}{Bartolomé, F.}
\newblock \bibinfo{title}{Chapter 1 - magnetic relaxation of lanthanide-based
  molecular magnets}.
\newblock vol.~\bibinfo{volume}{26} of \emph{\bibinfo{series}{Handbook of
  Magnetic Materials}}, \bibinfo{pages}{1--289} (\bibinfo{publisher}{Elsevier},
  \bibinfo{year}{2017}).
\newblock
  \urlprefix\url{https://www.sciencedirect.com/science/article/pii/S1567271917300033}.

\bibitem{jensen1991rare}
\bibinfo{author}{Jensen, J.}, \bibinfo{author}{Mackintosh, A.} \&
  \bibinfo{author}{Mackintosh, B.}
\newblock \emph{\bibinfo{title}{Rare Earth Magnetism: Structures and
  Excitations}}.
\newblock International Series of Monographs on Physics
  (\bibinfo{publisher}{Clarendon Press}, \bibinfo{year}{1991}).
\newblock \urlprefix\url{https://books.google.com/books?id=LbTvAAAAMAAJ}.

\bibitem{PhysRevB.103.184419.Zhang.spin.H}
\bibinfo{author}{Zhang, Z.} \emph{et~al.}
\newblock \bibinfo{title}{Effective magnetic hamiltonian at finite temperatures
  for rare-earth chalcogenides}.
\newblock \emph{\bibinfo{journal}{Phys. Rev. B}}
  \textbf{\bibinfo{volume}{103}}, \bibinfo{pages}{184419}
  (\bibinfo{year}{2021}).
\newblock \urlprefix\url{https://link.aps.org/doi/10.1103/PhysRevB.103.184419}.

\bibitem{Callen}
\bibinfo{author}{Callen, E.} \& \bibinfo{author}{Callen, H.~B.}
\newblock \bibinfo{title}{Magnetostriction, forced magnetostriction, and
  anomalous thermal expansion in ferromagnets}.
\newblock \emph{\bibinfo{journal}{Phys. Rev.}} \textbf{\bibinfo{volume}{139}},
  \bibinfo{pages}{A455--A471} (\bibinfo{year}{1965}).
\newblock \urlprefix\url{https://link.aps.org/doi/10.1103/PhysRev.139.A455}.

\bibitem{PhysRevB.61.9130}
\bibinfo{author}{Lovesey, S.~W.} \& \bibinfo{author}{Staub, U.}
\newblock \bibinfo{title}{Magnetoelastic model for the relaxation of lanthanide
  ions in
  ${\mathrm{yba}}_{2}{\mathrm{cu}}_{3}{\mathrm{o}}_{7\ensuremath{-}\ensuremath{\delta}}$
  observed by neutron scattering}.
\newblock \emph{\bibinfo{journal}{Phys. Rev. B}} \textbf{\bibinfo{volume}{61}},
  \bibinfo{pages}{9130--9139} (\bibinfo{year}{2000}).
\newblock \urlprefix\url{https://link.aps.org/doi/10.1103/PhysRevB.61.9130}.

\end{thebibliography}


\begin{thebibliography}{10}
\expandafter\ifx\csname url\endcsname\relax
  \def\url#1{\texttt{#1}}\fi
\expandafter\ifx\csname urlprefix\endcsname\relax\def\urlprefix{URL }\fi
\providecommand{\bibinfo}[2]{#2}
\providecommand{\eprint}[2][]{\url{#2}}

\bibitem{PhysRevB.50.17953}
\bibinfo{author}{Bl\"ochl, P.~E.}
\newblock \bibinfo{title}{Projector augmented-wave method}.
\newblock \emph{\bibinfo{journal}{Phys. Rev. B}} \textbf{\bibinfo{volume}{50}},
  \bibinfo{pages}{17953--17979} (\bibinfo{year}{1994}).
\newblock \urlprefix\url{https://link.aps.org/doi/10.1103/PhysRevB.50.17953}.

\bibitem{KRESSE199615}
\bibinfo{author}{Kresse, G.} \& \bibinfo{author}{Furthmüller, J.}
\newblock \bibinfo{title}{Efficiency of ab-initio total energy calculations for
  metals and semiconductors using a plane-wave basis set}.
\newblock \emph{\bibinfo{journal}{Computational Materials Science}}
  \textbf{\bibinfo{volume}{6}}, \bibinfo{pages}{15--50} (\bibinfo{year}{1996}).
\newblock
  \urlprefix\url{https://www.sciencedirect.com/science/article/pii/0927025696000080}.

\bibitem{PhysRevB.54.11169}
\bibinfo{author}{Kresse, G.} \& \bibinfo{author}{Furthm\"uller, J.}
\newblock \bibinfo{title}{Efficient iterative schemes for ab initio
  total-energy calculations using a plane-wave basis set}.
\newblock \emph{\bibinfo{journal}{Phys. Rev. B}} \textbf{\bibinfo{volume}{54}},
  \bibinfo{pages}{11169--11186} (\bibinfo{year}{1996}).
\newblock \urlprefix\url{https://link.aps.org/doi/10.1103/PhysRevB.54.11169}.

\bibitem{PhysRevB.59.1758}
\bibinfo{author}{Kresse, G.} \& \bibinfo{author}{Joubert, D.}
\newblock \bibinfo{title}{From ultrasoft pseudopotentials to the projector
  augmented-wave method}.
\newblock \emph{\bibinfo{journal}{Phys. Rev. B}} \textbf{\bibinfo{volume}{59}},
  \bibinfo{pages}{1758--1775} (\bibinfo{year}{1999}).
\newblock \urlprefix\url{https://link.aps.org/doi/10.1103/PhysRevB.59.1758}.

\bibitem{PhysRevLett.77.3865}
\bibinfo{author}{Perdew, J.~P.}, \bibinfo{author}{Burke, K.} \&
  \bibinfo{author}{Ernzerhof, M.}
\newblock \bibinfo{title}{Generalized gradient approximation made simple}.
\newblock \emph{\bibinfo{journal}{Phys. Rev. Lett.}}
  \textbf{\bibinfo{volume}{77}}, \bibinfo{pages}{3865--3868}
  (\bibinfo{year}{1996}).
\newblock \urlprefix\url{https://link.aps.org/doi/10.1103/PhysRevLett.77.3865}.

\bibitem{togo2015first}
\bibinfo{author}{Togo, A.} \& \bibinfo{author}{Tanaka, I.}
\newblock \bibinfo{title}{First principles phonon calculations in materials
  science}.
\newblock \emph{\bibinfo{journal}{Scripta Materialia}}
  \textbf{\bibinfo{volume}{108}}, \bibinfo{pages}{1--5} (\bibinfo{year}{2015}).

\bibitem{PhysRevB.57.1505}
\bibinfo{author}{Dudarev, S.~L.}, \bibinfo{author}{Botton, G.~A.},
  \bibinfo{author}{Savrasov, S.~Y.}, \bibinfo{author}{Humphreys, C.~J.} \&
  \bibinfo{author}{Sutton, A.~P.}
\newblock \bibinfo{title}{Electron-energy-loss spectra and the structural
  stability of nickel oxide: An lsda+u study}.
\newblock \emph{\bibinfo{journal}{Phys. Rev. B}} \textbf{\bibinfo{volume}{57}},
  \bibinfo{pages}{1505--1509} (\bibinfo{year}{1998}).
\newblock \urlprefix\url{https://link.aps.org/doi/10.1103/PhysRevB.57.1505}.

\bibitem{doi:10.1063/1.3382344}
\bibinfo{author}{Grimme, S.}, \bibinfo{author}{Antony, J.},
  \bibinfo{author}{Ehrlich, S.} \& \bibinfo{author}{Krieg, H.}
\newblock \bibinfo{title}{A consistent and accurate ab initio parametrization
  of density functional dispersion correction (dft-d) for the 94 elements
  h-pu}.
\newblock \emph{\bibinfo{journal}{The Journal of Chemical Physics}}
  \textbf{\bibinfo{volume}{132}}, \bibinfo{pages}{154104}
  (\bibinfo{year}{2010}).
\newblock \urlprefix\url{https://doi.org/10.1063/1.3382344}.
\newblock \eprint{https://doi.org/10.1063/1.3382344}.

\bibitem{pai2021nearlyresonant}
\bibinfo{author}{Pai, Y.-Y.} \emph{et~al.}
\newblock \bibinfo{title}{Mesoscale interplay between phonons and crystal
  electric field excitations in quantum spin liquid candidate csybse2}.
\newblock \emph{\bibinfo{journal}{J. Mater. Chem. C}} \bibinfo{pages}{--}
  (\bibinfo{year}{2022}).
\newblock \urlprefix\url{http://dx.doi.org/10.1039/D1TC05934C}.

\bibitem{pyMC}
\bibinfo{author}{Salvatier, J.}, \bibinfo{author}{Wiecki, T.~V.} \&
  \bibinfo{author}{Fonnesbeck, C.}
\newblock \bibinfo{title}{Probabilistic programming in python using pymc3}.
\newblock \emph{\bibinfo{journal}{PeerJ Computer Science}}
  \textbf{\bibinfo{volume}{2}}, \bibinfo{pages}{e55} (\bibinfo{year}{2016}).
\newblock \urlprefix\url{https://doi.org/10.7717/peerj-cs.55}.

\bibitem{jensen1991rare}
\bibinfo{author}{Jensen, J.}, \bibinfo{author}{Mackintosh, A.} \&
  \bibinfo{author}{Mackintosh, B.}
\newblock \emph{\bibinfo{title}{Rare Earth Magnetism: Structures and
  Excitations}}.
\newblock International Series of Monographs on Physics
  (\bibinfo{publisher}{Clarendon Press}, \bibinfo{year}{1991}).
\newblock \urlprefix\url{https://books.google.com/books?id=LbTvAAAAMAAJ}.

\bibitem{NaYbSe2PRB_CEF}
\bibinfo{author}{Zhang, Z.} \emph{et~al.}
\newblock \bibinfo{title}{Crystalline electric field excitations in the quantum
  spin liquid candidate ${{NaYbSe}}_{2}$}.
\newblock \emph{\bibinfo{journal}{Phys. Rev. B}}
  \textbf{\bibinfo{volume}{103}}, \bibinfo{pages}{035144}
  (\bibinfo{year}{2021}).
\newblock \urlprefix\url{https://link.aps.org/doi/10.1103/PhysRevB.103.035144}.

\bibitem{Schmidt_2021}
\bibinfo{author}{Schmidt, B.}, \bibinfo{author}{Sichelschmidt, J.},
  \bibinfo{author}{Ranjith, K.~M.}, \bibinfo{author}{Doert, T.} \&
  \bibinfo{author}{Baenitz, M.}
\newblock \bibinfo{title}{Yb delafossites: Unique exchange frustration of $4f$
  spin-$\frac{1}{2}$ moments on a perfect triangular lattice}.
\newblock \emph{\bibinfo{journal}{Phys. Rev. B}}
  \textbf{\bibinfo{volume}{103}}, \bibinfo{pages}{214445}
  (\bibinfo{year}{2021}).
\newblock \urlprefix\url{https://link.aps.org/doi/10.1103/PhysRevB.103.214445}.

\bibitem{pocs2021systematic}
\bibinfo{author}{Pocs, C.~A.} \emph{et~al.}
\newblock \bibinfo{title}{Systematic fitting of crystal-field levels and
  accurate extraction of quantum magnetic models in triangular-lattice
  delafossites} (\bibinfo{year}{2021}).

\bibitem{scheie2021witnessing}
\bibinfo{author}{Scheie, A.~O.} \emph{et~al.}
\newblock \bibinfo{title}{Witnessing quantum criticality and entanglement in
  the triangular antiferromagnet kybse$_2$} (\bibinfo{year}{2021}).
\newblock \eprint{2109.11527}.

\end{thebibliography}

\end{document}


\title{Supplementary Information for Phonon Chirality Induced by Vibronic Orbital Coupling}

\author{Yun-Yi Pai}
\email{yunyip@ornl.gov}
\address{Materials Science and Technology Division, Oak Ridge National Laboratory, Oak Ridge, TN 37831, USA}
\address{Quantum Science Center, Oak Ridge, Tennessee 37831, USA}

\author{Claire E. Marvinney}
\address{Materials Science and Technology Division, Oak Ridge National Laboratory, Oak Ridge, TN 37831, USA}
\address{Quantum Science Center, Oak Ridge, Tennessee 37831, USA}

\author{Liangbo Liang}
\address{Center for Nanophase Materials Sciences, Oak Ridge National Laboratory, Oak Ridge, TN 37831, USA}

\author{Ganesh Pokharel}
\address{Materials Department and California Nanosystems Institute, University of California Santa Barbara, Santa Barbara, CA 93106, USA}

\author{Jie Xing}
\address{Materials Science and Technology Division, Oak Ridge National Laboratory, Oak Ridge, TN 37831, USA}

\author{Haoxiang Li}
\address{Materials Science and Technology Division, Oak Ridge National Laboratory, Oak Ridge, TN 37831, USA}

\author{Xun Li}
\address{Materials Science and Technology Division, Oak Ridge National Laboratory, Oak Ridge, TN 37831, USA}

\author{Michael Chilcote}
\address{Materials Science and Technology Division, Oak Ridge National Laboratory, Oak Ridge, TN 37831, USA}
\address{Quantum Science Center, Oak Ridge, Tennessee 37831, USA}

\author{Matthew Brahlek}
\address{Materials Science and Technology Division, Oak Ridge National Laboratory, Oak Ridge, TN 37831, USA}
\address{Quantum Science Center, Oak Ridge, Tennessee 37831, USA}

\author{Lucas Lindsay}
\address{Materials Science and Technology Division, Oak Ridge National Laboratory, Oak Ridge, TN 37831, USA}

\author{Hu Miao}
\address{Materials Science and Technology Division, Oak Ridge National Laboratory, Oak Ridge, TN 37831, USA}

\author{Athena S. Sefat}
\address{Materials Science and Technology Division, Oak Ridge National Laboratory, Oak Ridge, TN 37831, USA}

\author{David Parker}
\address{Materials Science and Technology Division, Oak Ridge National Laboratory, Oak Ridge, TN 37831, USA}

\author{Stephen D. Wilson}
\address{Materials Department and California Nanosystems Institute, University of California Santa Barbara, Santa Barbara, CA 93106, USA}

\author{Benjamin J. Lawrie}
\email{lawriebj@ornl.gov}
\address{Materials Science and Technology Division, Oak Ridge National Laboratory, Oak Ridge, TN 37831, USA}
\address{Quantum Science Center, Oak Ridge, Tennessee 37831, USA}

\date{\today}
\pacs{}
\maketitle
\tableofcontents

\section{Phonon calculation details}
To obtain phonon dispersion in NaYbSe$_2$, density functional theory (DFT) calculations were performed based on the projector augmented wave method (PAW) as implemented in the Vienna Ab-initio Simulation Package (VASP) \cite{PhysRevB.50.17953, KRESSE199615, PhysRevB.54.11169, PhysRevB.59.1758}. The generalized gradient approximation, parameterized by Perdew, Burke, and Ernzerhof (PBE) \cite{PhysRevLett.77.3865} was used for exchange-correlations. A 520 eV kinetic energy cutoff in the plane-wave expansion and energy convergence criteria of 10$^{-6}$~eV were employed. Ionic relaxations were performed until Hellmann-Feynman forces converged to 10$^{-4}$~meV/\AA. The structure was relaxed with a $\Gamma$-centered 15$\times$15$\times$3 k-mesh. The harmonic interatomic force constants (IFCs) were calculated using the finite displacement method implemented in the phonopy package \cite{togo2015first} in a 3$\times$3$\times$1 supercell with $\Gamma$-centered 5$\times$5$\times$3 k-meshes. The DFT+U method \cite{PhysRevB.57.1505} was used to include the Coulomb correlations with U$_{\text{eff}}$=6~eV~\cite{PhysRevB.57.1505} for Yb atoms. Van der Waals interactions were taken into account via DFT-D3 method \cite{doi:10.1063/1.3382344}.

\begin{figure}[htbp]
\centering
\includegraphics[width=0.8\linewidth]{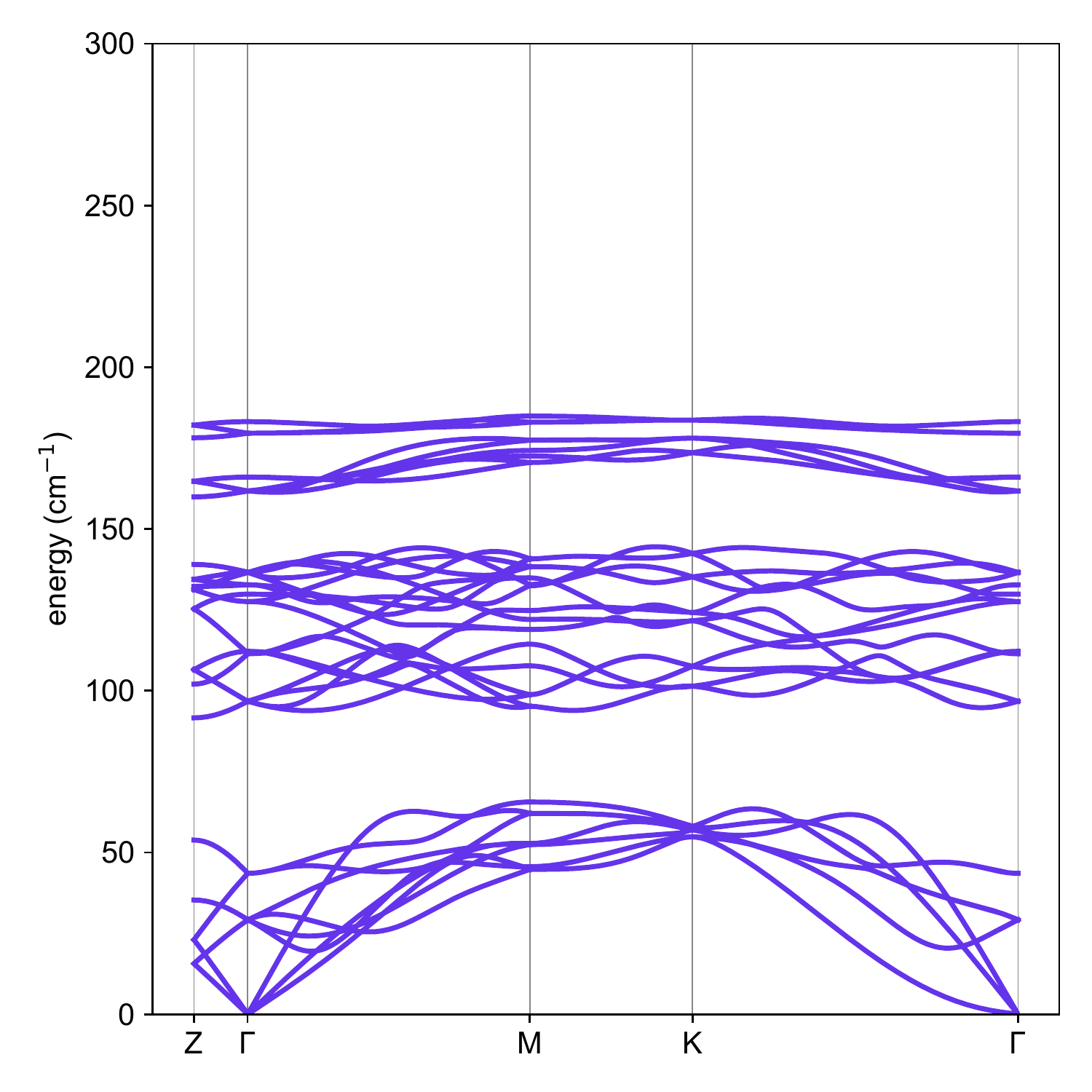}
\caption{Calculated phonon dispersion.}
\label{fig:phonon_DFT}
\end{figure}

\clearpage

\section{Temperature Dependence for Higher Energy Band}
\begin{figure}[htbp]
\centering
\includegraphics[width=1.0\linewidth]{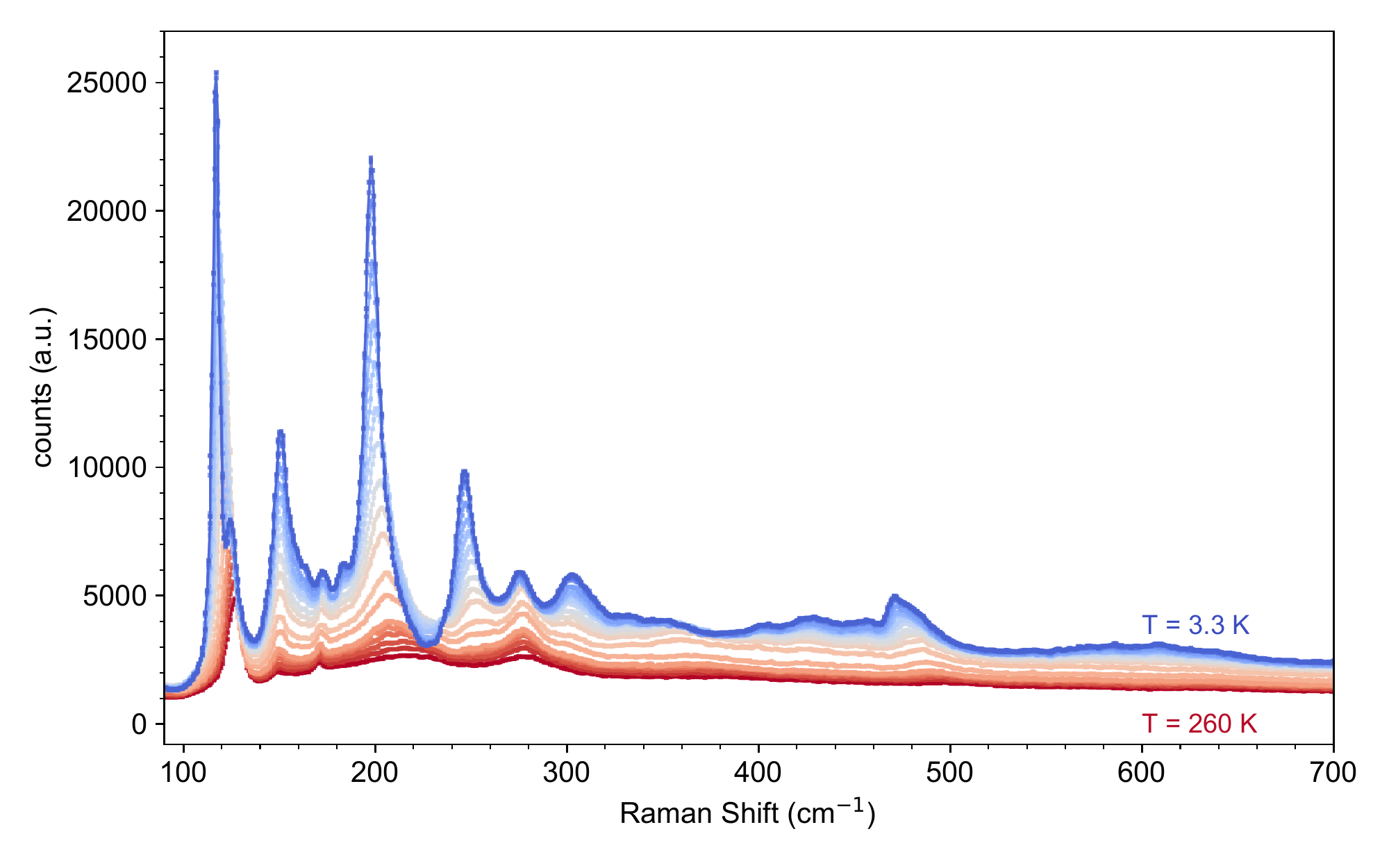}
\caption{Raman spectra as function of temperature from $T =$ 3.3 K to $T =$ 270 K.}
\label{fig:full}
\end{figure}

Figure \ref{fig:full} shows temperature-dependent unpolarized Raman spectra taken from $T =$ 3.3 K to $T =$ 270 K. The spectra were taken with Semrock dichroic and longpass filters with cutoff at 90 cm$^{-1}$ instead of a set of volume Bragg gratings. Similar to CsYbSe$_2$ \cite{pai2021nearlyresonant}, a multitude of possible combination modes (e.g., possible CEF1 + CEF2) -- behavior of resonant Raman excitation -- is present.

\begin{figure}[htbp]
\centering
\includegraphics[width=1.0\linewidth]{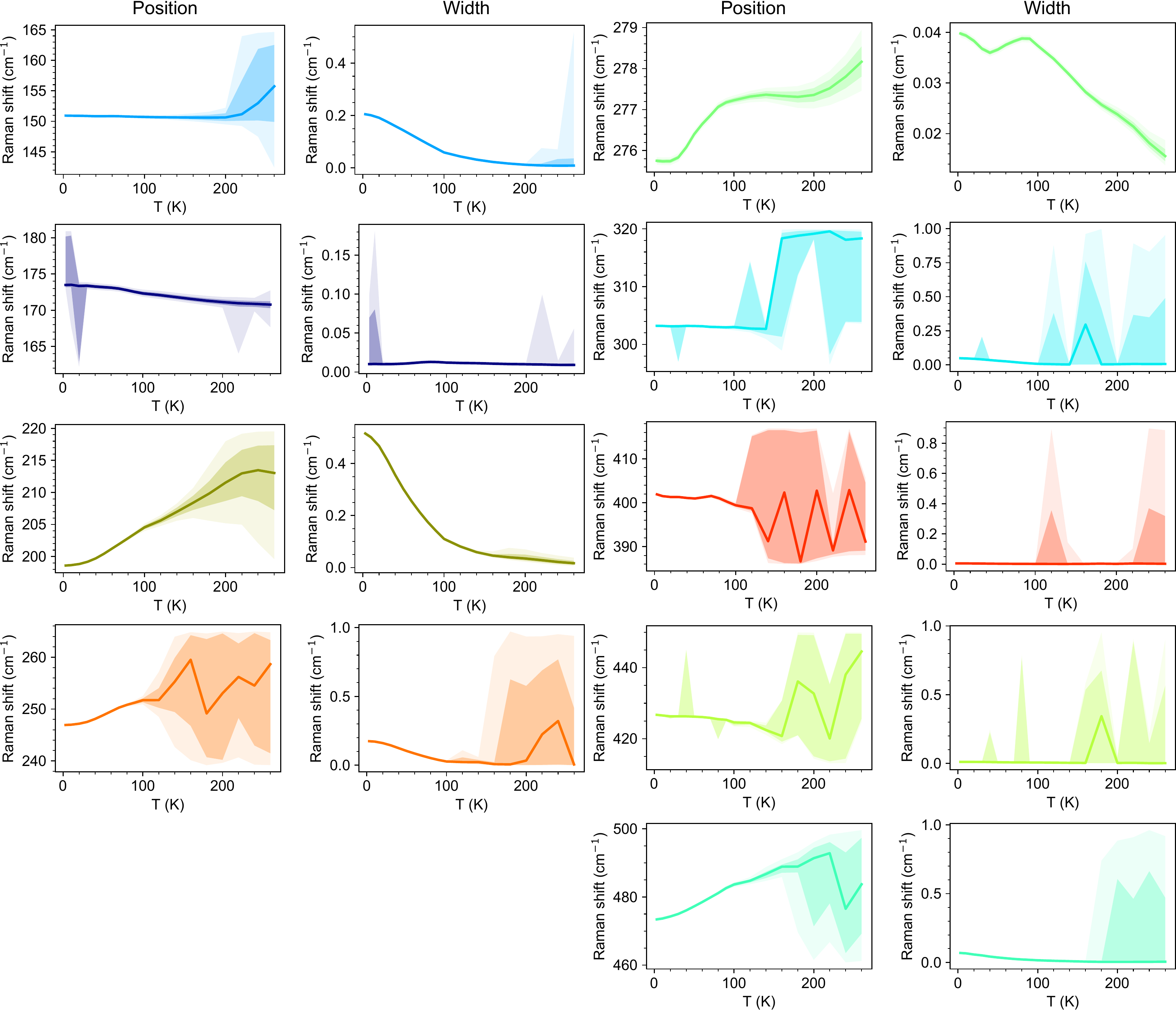}
\caption{Peak postitions extracted from Figure \ref{fig:full}.}
\label{fig:peak_pos}
\end{figure}

Figure \ref{fig:peak_pos} shows the peak positions extracted from Figure \ref{fig:full} using Bayesian inference \cite{pyMC}. In each subplot, the trace is the median of the posterior distribution of the peak position or peak width, and the two shades of different opacities are corresponding to 1$\sigma$ (more opaque) and 2$\sigma$ (more transparent) bands of the posterior distribution.  

\clearpage

\section{Temperature Dependence for XX and XY polarization}
\begin{figure}[htbp]
\centering
\includegraphics[width=1.0\linewidth]{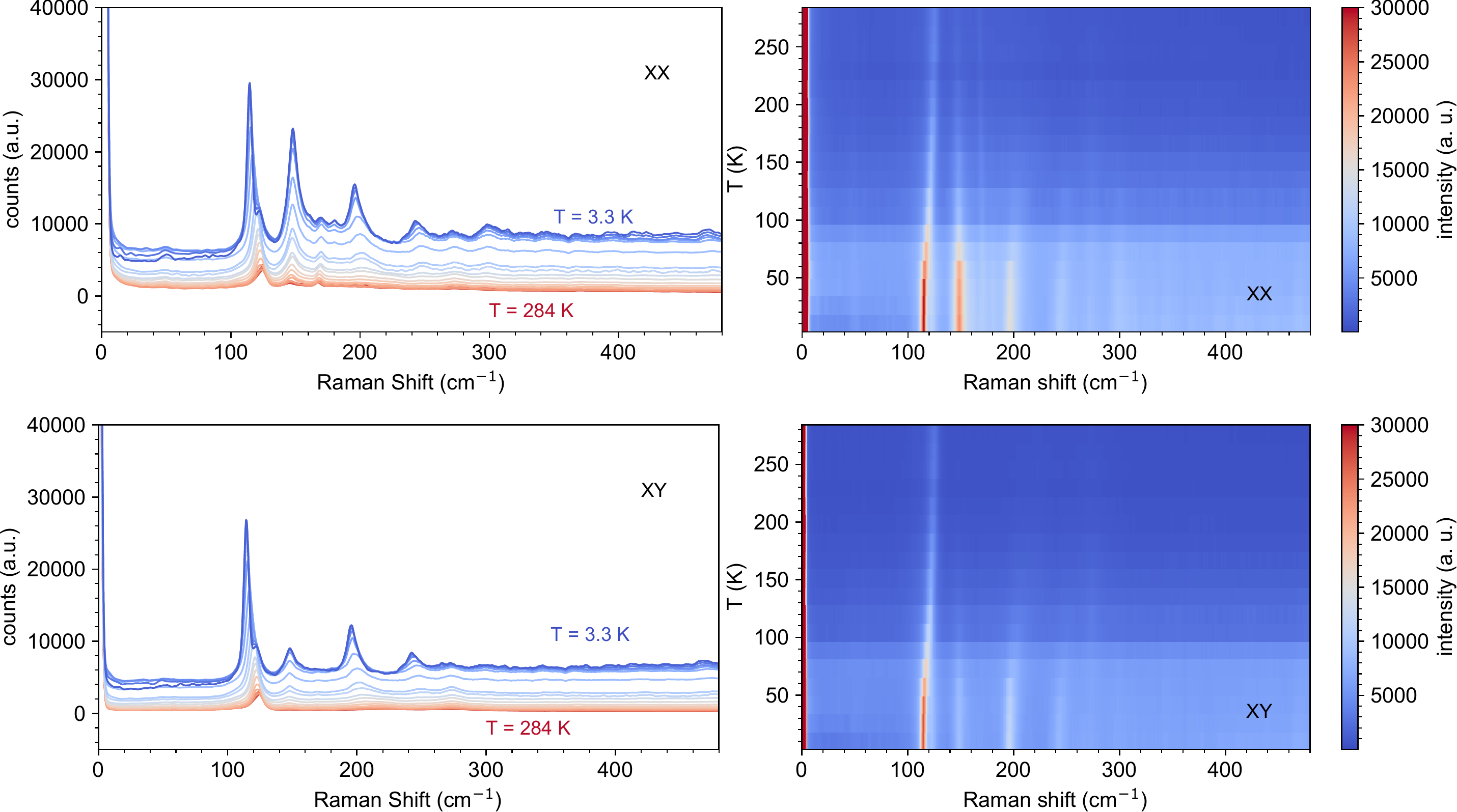}
\caption{Raman spectra taken from $T =$ 3.3 K to $T =$ 270 K with polarization configuration $XX$ and $XY$.}
\label{fig:temp_pol}
\end{figure}

Figure \ref{fig:temp_pol} shows temperature-dependent Raman spectra taken from $T =$ 3.3 K to $T =$ 270 K with polarization configuration $XX$ and $XY$.  

\clearpage

\section{Full Magneto-Raman for both N\lowercase{a}Y\lowercase{b}S\lowercase{e}$_2$ and C\lowercase{s}Y\lowercase{b}S\lowercase{e}$_2$ at $T$ = 4K}
\begin{figure}[htbp]
\centering
\includegraphics[width=1.0\linewidth]{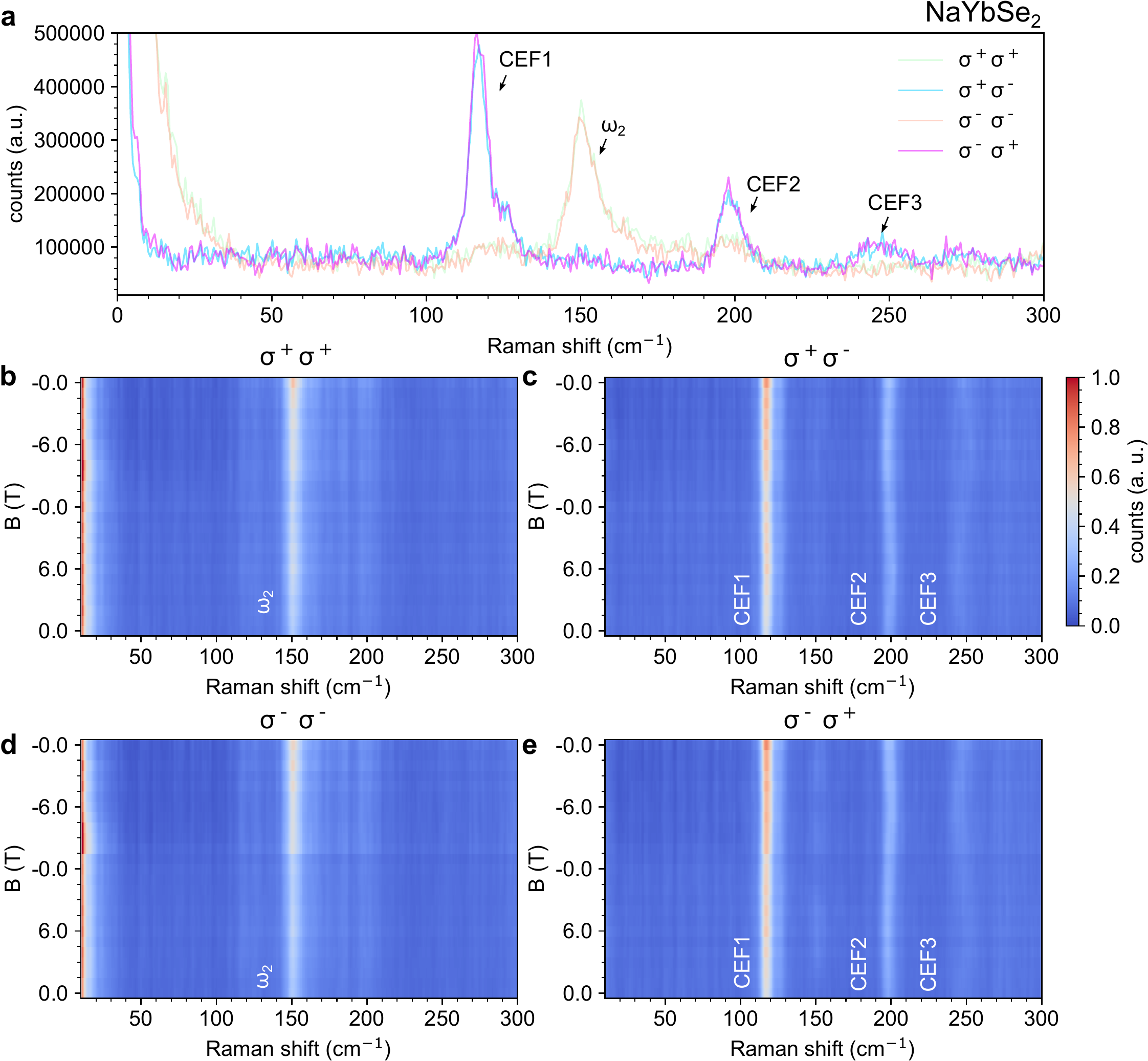}
\caption{Helicity-resolved magnetic field dependence of CEFs and $\omega$ for NaYbSe$_2$ at $T = 4$ K. (a) at 0T, (b) ($\sigma^{+}, \sigma^{+}$). (c) ($\sigma^{+}, \sigma^{-}$). (d) ($\sigma^{-}, \sigma^{-}$). (e) ($\sigma^{-}, \sigma^{+}$).  }
\label{fig:magnetoNa}
\end{figure}

\begin{figure}[htbp]
\centering
\includegraphics[width=1.0\linewidth]{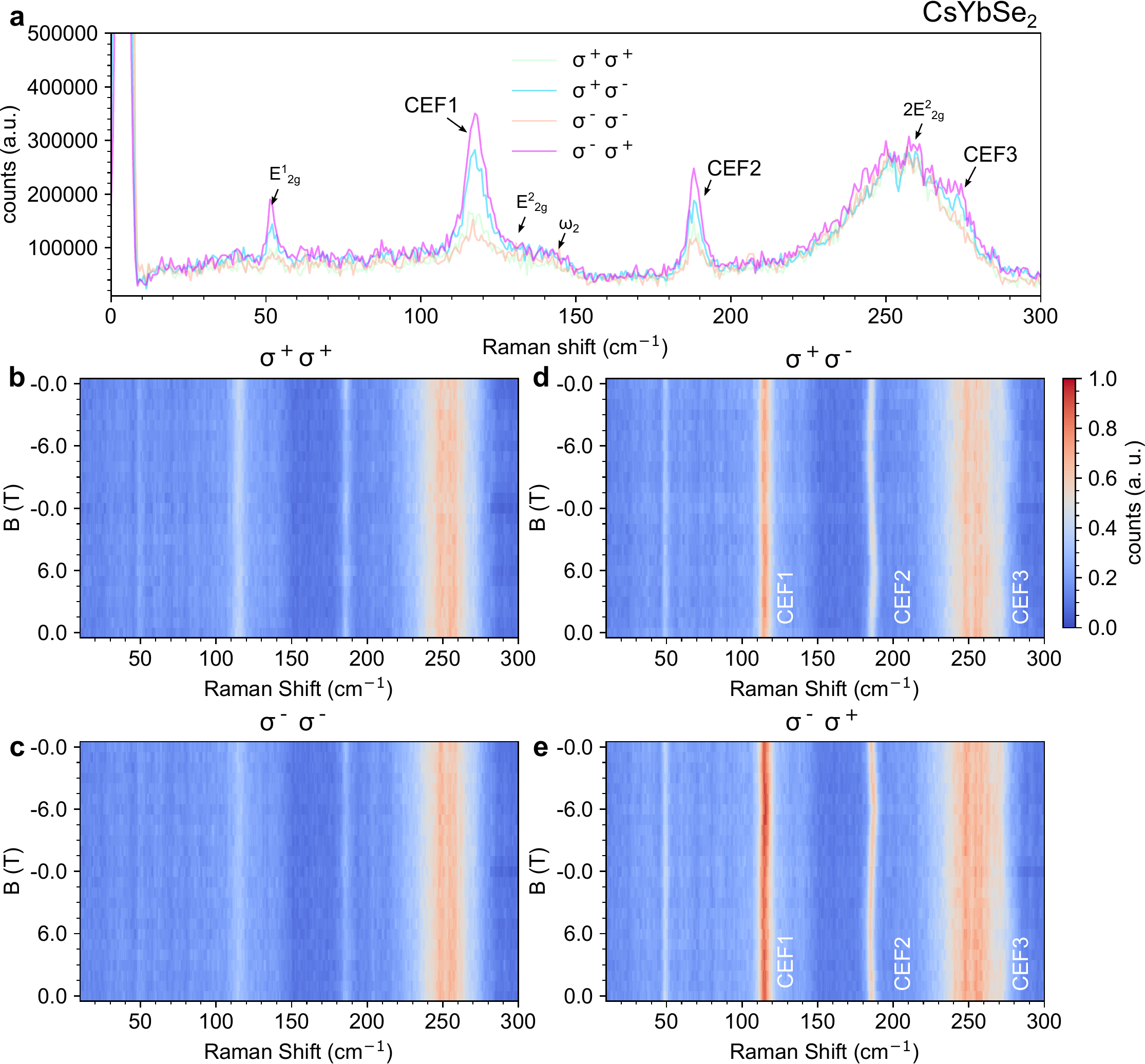}
\caption{Helicity-resolved magnetic field dependence of CEFs and $\omega$ for CsYbSe$_2$ at $T = 4$ K. (a) at 0T,  (b) ($\sigma^{+}, \sigma^{+}$). (c) ($\sigma^{+}, \sigma^{-}$). (d) ($\sigma^{-}, \sigma^{-}$). (e) ($\sigma^{-}, \sigma^{+}$).  }
\label{fig:magnetoCs}
\end{figure}

Figure \ref{fig:magnetoNa} shows the original magneto-Raman spectra for NaYbSe$_2$ and Figure \ref{fig:magnetoCs} for CsYbSe$_2$. The magnetic field dependence given polarization configuration for the CEF modes are similar for NaYbSe$_2$ and CsYbSe$_2$, i.e., (i) with ($\sigma_{\text{incident}}$, $\sigma_{\text{scatter}}$) = ($\sigma^+$, $\sigma^-$) configuration, the CEF1 and CEF2 shift toward higher energy in positive field, and CEF3 toward lower energy and (ii) the trends are inverted with inverted polarization configurations or inverted field. However, the $\omega$ is observed only in a subset of spatial locations \cite{pai2021nearlyresonant}, and with much weaker intensity, likely due to larger detuning between the $E_{\text{2g}}^2$ and CEF1 in CsYbSe$_2$. Small residual peaks that violate the proposed selection rules are present in both datasets: for instance, a small $\omega$ peak is present in Figure \ref{fig:magnetoNa} (c) and Figure \ref{fig:magnetoNa} (e) due to less than ideal polarization contrast in the optics train. For CsYbSe$_2$, the extinction ratio of the selection rules are not as high as that of NaYbSe$_2$, suggesting that the selection rules (and hence the assignment of the states) proposed in the text may only apply to NaYbSe$_2$.

\clearpage
\section{Assignment of the CEF to eigenstates for N\lowercase{a}Y\lowercase{b}S\lowercase{e}$_2$}

The CEF Hamiltonian for NaYbSe$_2$ within the point charge approximation \cite{jensen1991rare,NaYbSe2PRB_CEF,Schmidt_2021,pocs2021systematic} that describes the ground state manifold $J = 7/2$ is: 
\begin{equation}
\begin{split}
    H_{\text{CEF}} = &B_2^0 \textbf{O}_2^0 + B_4^0 \textbf{O}_4^0 + B_4^3 \textbf{O}_4^3 \\
    + &B_6^0 \textbf{O}_6^0 + B_6^3 \textbf{O}_6^3 + B_6^6 \textbf{O}_6^6\
\end{split}
\end{equation}

The expected helicity dependence of the CEF excitations follows from the selection rule bridging the two relevant states. The eigenstates for the energy levels described in equation 1 are: 
\begin{equation}
\begin{split}
    | \psi^+_{\text{1, 2, 3}} \rangle = &-\alpha e^{i \phi_\alpha}  | \frac{7}{2}, \frac{7}{2} \rangle  + \beta | \frac{7}{2}, \frac{1}{2} \rangle + \gamma e^{-i \phi_\gamma}  | \frac{7}{2}, -\frac{5}{2} \rangle\\
    | \psi^-_{\text{1, 2, 3}} \rangle = &\alpha e^{i \phi_\alpha}  | \frac{7}{2}, -\frac{7}{2} \rangle  + \beta | \frac{7}{2}, -\frac{1}{2} \rangle - \gamma e^{i \phi_\gamma}  | \frac{7}{2}, \frac{5}{2} \rangle\\
\end{split}
\end{equation}
and 
\begin{equation}
\begin{split}
    | \psi^+_0 \rangle = & | \frac{7}{2}, \frac{3}{2} \rangle  \\
    | \psi^-_0 \rangle = & | \frac{7}{2}, -\frac{3}{2} \rangle \\
\end{split}
\end{equation}
where $\alpha$, $\beta$, $\gamma$, $\phi_\alpha$, $\phi_\beta$, and $\phi_\gamma$ $\in \mathbb{R}$ are determined by the CEF parameters $B_2^0$, $B_4^0$, $B_4^3$, $B_6^0$, $B_6^3$ and $B_6^6$. Note that the levels are doubly degenerate. 

Without further constraints, the 6 parameters have enough degrees of freedom to fit experimentally observed energy levels. More than one set of CEF parameters that locally minimizes the error may exist and the order of the eigenstates may not be the same across those sets. For example, the subspace spanned by this special singlet pair with single angular momentum eigenstate $ | \psi^{\pm}_0 \rangle$ is assigned to CEF1 in Zhang et al. \cite{NaYbSe2PRB_CEF} and CEF2 in Scheie et al. \cite{scheie2021witnessing}. Schimidt et al. \cite{Schmidt_2021} pointed out that they cannot be the ground state due to observed in-plane field dependence at low temperatures \cite{Schmidt_2021}. 

However, the observed CEFs are dominantly in cross circular channel, i.e., ($\sigma^+, \sigma^-$) and ($\sigma^-, \sigma^+$) ($J_+J_+$ and $J_-J_-$). Note that the sign for the scattered light is reversed due to the backscattering geometry. On the other hand, CEFs have reduced to negligible intensity in co-circular channel, ($\sigma^+, \sigma^+$) and ($\sigma^-, \sigma^-$) (equivalently $J_+J_-$ and $J_-J_+$).  

The observation above forces the singlet eigenspace $ | \psi^{\pm}_0 \rangle$ to be the ground state. \textbf{Mathematical Argument: } We argue the statement above by contradiction. First, without loss of generality, we consider the following eigenstates assignment: C

$|\psi^{\pm}_{\text{1, 2, 3}}$ are 0$\pm$, 2$\pm$ and 3$\pm$ and $|\psi^{\pm}_{\text{0}}$ are $1\pm$

as show in the diagram on the right: the singlets are assigned to $1+/1-$. The space spanned by $0+$, $2+$, and $3+$ is a 3-dimensional space. Note that by definition, the state $0+$, $2+$, and $3+$ are orthogonal to each other. The set orthogonality conditions, i.e., $\langle 0+|2+ \rangle=0$, $\langle 0+|3+ \rangle=0$, $\langle 2+|3+ \rangle=0$, are a set of 3 bilinear equations. They have a simple geometry interpretation: While the first vector is completely free, once the first vector is assigned, the 3 orthogonality conditions set the other 2 vectors, up to a permutation and normalization. Now, since the matrix elements bridged by $J_+J_-$ and $J_-J_+$ have to be zero, it means $\langle 0+| J_+J_- | 2+ \rangle=0$, $\langle 0+| J_+J_- | 3+ \rangle=0$, $\langle 0+| J_-J_+ | 2+ \rangle=0$, $\langle 0+| J_-J_+ | 3+ \rangle=0$. But because $J_+J_- \neq I$ (identity) and $J_-J_+ \neq I$, the 4 conditions above impose 4 new sets of bilinear equations for the system to satisfy. The solution therefore does not exist (A line that is orthogonal to a set of 4 skew lines in 3D space does not exists) $\Rightarrow\mspace{-4mu}\Leftarrow$.

\newpage

\subsection{Fitted parameters for NaYbSe$_2$}
To fit the experimental data, we used a general non-linear model normalization procedure in Mathematica. To start, a general set of eigenvalues and eigenvectors as a function of the CEF parameters is obtained by direct diagonalization of the CEF Hamiltonian. The eigenvectors are sorted based on their eigenvalues for each test parameter space. Then the cost function is defined by the sum of the squared errors between transitions and the data, with both inter-branch and intra-branch transitions are considered. The selection rules are added as a term in the cost function when necessary. The final CEF parameters is the set that minimized the said cost function.   

Table \ref{tab:table-na-eigen} is the obtained eigenstates for CsYbSe$_2$ and Table \ref{tab:na-sele} matrix elements between the eigenstates, acted by  $J_+J_+$, $J_-J_-$, $J_+J_-$, $J_-J_+$.  Note that the matrix elements in the $J_+J_-$ and $J_-J_+$ channels are all $0$, consistent with experimental observations. 
\begin{table}[htbp]
\begin{tabular}{c | cccc cccc}
 state\textbackslash $m_j$  & -7/2 & -5/2 & -3/2 & -1/2 & 1/2 & 3/2 & 5/2 & 7/2 \\ \hline
0- & 0    & 0    & 1    & 0    & 0   & 0   & 0   & 0   \\
0+ & 0    & 0    & 0    & 0    & 1   & 0   & 0   & 0   \\
1- & 0.725952   & 0    & 0    & 0.0000367007   & 0   & 0   & 0.687745  & 0   \\
1+ & 0    & 0.687745    & 0    & 0    & -0.0000367007   & 0   & 0   & 0.725953   \\
2- & -0.687745    & 0    & 0    & 0.0000163497    & 0   & 0   & 0.725952   & 0   \\
2+ & 0    & -0.725953    & 0    & 0    & 0.0000163497  & 0   & 0   & 0.687745  \\
3- & 0.0000153987    & 0    & 0    & -1.    & 0   & 0   & 0.0000371098   & 0   \\
3+ & 0    & 0.0000371099    & 0    & 0    & 1.  & 0   & 0   & 0.000015398  
\end{tabular}
\caption{\label{tab:table-na-eigen}Eigenstates for NaYbSe$_2$}
\end{table}

\begin{table}[htbp]
\begin{tabular}{c | cccc}
combination\textbackslash $O_{op}$ & $J_+J_+$ & $J_-J_-$ & $J_+J_-$ & $J_-J_+$ \\\hline
 $|| \langle 1- | O_{op} | 0- \rangle ||^2$  & 0        &  44.2686 & 0   & 0 \\
 $|| \langle 1+ | O_{op} | 0- \rangle ||^2$  & 0        & 3.23267$\times 10^{-7}$
       & 0  & 0 \\
 $|| \langle 1+ | O_{op} | 0+ \rangle ||^2$  &   44.2686

 & 0       & 0   & 0 \\
 $|| \langle 1- | O_{op} | 0+ \rangle ||^2$  &  3.23267$\times 10^{-7}$      & 0       & 0   &0 \\\hline

  $|| \langle 2- | O_{op} | 0- \rangle ||^2$  & 0  & 39.7314      & 0& 0 \\
 $|| \langle 2+ | O_{op} | 0- \rangle ||^2$  &  0 & 6.41549$\times 10^{-8}$      & 0   & 0 \\
 $|| \langle 2+ | O_{op} | 0+ \rangle ||^2$  & 39.7314        & 0 & 0  &  0 \\
 $|| \langle 2- | O_{op} | 0+ \rangle ||^2$  &  6.41549$\times 10^{-8}$        & 0 & 0   & 0  \\\hline
 
 $|| \langle 3- | O_{op} | 0- \rangle ||^2$  & 0        & 1.9918$\times 10^{-8}$       &  0        &  0 \\
 $|| \langle 3+ | O_{op} | 0- \rangle ||^2$  & 0        & 240 & 0         & 0       \\
 $|| \langle 3+ | O_{op} | 0+ \rangle ||^2$  & 0        & 0       &  0        &  0 \\
 $|| \langle 3- | O_{op} | 0+ \rangle ||^2$  & 240  &  42.5604       & 0         & 0       
\end{tabular}
\caption{\label{tab:na-sele} Matrix element for  $O_{op}$ $ = J_+J_+$,  $J_-J_-$,  $J_+J_-$,  $J_-J_+$ between CEF eigenstates for NaYbSe$_2$ using eigenstates in Table \ref{tab:table-na-eigen}.}
\end{table}

\clearpage
Addtionally, we compare the eigenstates in  Zhang et al. \cite{NaYbSe2PRB_CEF}, which is listed in Table \ref{tab:table-zhang}. Table \ref{tab:table-zhang-sele} lists the matrix elements between the eigenstates, acted by  $J_+J_+$, $J_-J_-$, $J_+J_-$, $J_-J_+$. Not all of the matrix elements in the $J_+J_-$ and $J_-J_+$ channels are all $0$. For matrix elements larger than 50\% of matrix elements of those in $J_+J_+$, $J_-J_-$ are highlighted in \textcolor{red}{red}.

\begin{table}[htbp]
\begin{tabular}{c | cccc cccc}
 state\textbackslash $m_j$  & -7/2 & -5/2 & -3/2 & -1/2 & 1/2 & 3/2 & 5/2 & 7/2 \\ \hline
0- & 0    & 0.8019    & 0    & 0    &  -0.1368   & 0   & 0   &  -0.5693  \\
0+ & -0.5693    & 0    & 0    & 0.1368    & 0   & 0   & 0.8019   & 0   \\
1- & 0    & 0    &  -0.1396    & 0    & 0   &  -0.9902   & 0   & 0   \\
1+ & 0    & 0    &  +0.9902   & 0    & 0   &  -0.1396   & 0   & 0   \\
2- & 0    & -0.4716    & 0    & 0    &  -0.7307   & 0   & 0   &  -0.4887   \\
2+ &  -0.4887   & 0    & 0    &  0.7307    & 0   & 0   &  -0.4716   & 0   \\
3- & 0    &  -0.3524    & 0    & 0    &  0.6665  & 0   & 0   &  -0.6565   \\
3+ &  -0.6565    & 0    & 0    & -0.6665    & 0   & 0   &  -0.3524   & 0  
\end{tabular}
\caption{\label{tab:table-zhang}Eigenstates for NaYbSe$_2$ from Zhang et al. \cite{NaYbSe2PRB_CEF}}
\end{table}

\begin{table}[htbp]
\begin{tabular}{c | cccc}
combination\textbackslash $O_{op}$ & $J_+J_+$ & $J_-J_-$ & $J_+J_-$ & $J_-J_+$ \\\hline
 $|| \langle 1- | O_{op} | 0- \rangle ||^2$  & 0        &  29.8383 & 0   & 0 \\
 $|| \langle 1+ | O_{op} | 0- \rangle ||^2$  & 0        & 1.87727
       & 0  & 0 \\
 $|| \langle 1+ | O_{op} | 0+ \rangle ||^2$  &  29.8383
 & 0       & 0   & 0 \\
 $|| \langle 1- | O_{op} | 0+ \rangle ||^2$  &  1.87727        & 0       & 0   &0 \\\hline

  $|| \langle 2- | O_{op} | 0- \rangle ||^2$  & 0  & 0       & 0.809356   & 9.2338 \\
 $|| \langle 2+ | O_{op} | 0- \rangle ||^2$  &  76.1584  & 0       & 0   & 0 \\
 $|| \langle 2+ | O_{op} | 0+ \rangle ||^2$  & 0.       & 0 &  9.2338   &  0.809356 \\
 $|| \langle 2- | O_{op} | 0+ \rangle ||^2$  & 0        & 76.1584 & 0   & 0  \\\hline
 
 $|| \langle 3- | O_{op} | 0- \rangle ||^2$  & 0        & 0       &  0.673621        &  \textcolor{red}{22.6455} \\
 $|| \langle 3+ | O_{op} | 0- \rangle ||^2$  & 42.5604        & 0 & 0         & 0       \\
 $|| \langle 3+ | O_{op} | 0+ \rangle ||^2$  & 0        & 0       &  \textcolor{red}{22.6455}      & 0.673621 \\
 $|| \langle 3- | O_{op} | 0+ \rangle ||^2$  & 0  &  42.5604       & 0         & 0       
\end{tabular}
\caption{\label{tab:table-zhang-sele}Matrix element for  $O_{op}$ $ = J_+J_+$,  $J_-J_-$,  $J_+J_-$,  $J_-J_+$ between CEF eigenstates for NaYbSe$_2$ using eigenstates from Zhang et al. \cite{NaYbSe2PRB_CEF} }
\end{table}

\clearpage

\subsection{Eigenstates for CsYbSe$_2$}
For CsYbSe$_2$, Table \ref{tab:table-cs-eigen} is the obtained eigenstates for CsYbSe$_2$ and Table \ref{tab:cs-sele} matrix elements between the eigenstates, acted by  $J_+J_+$, $J_-J_-$, $J_+J_-$, $J_-J_+$. For the channel $J_+J_-$, $J_-J_+$, matrix elements larger than 50\% of that in  $J_+J_+$, $J_-J_-$ are highlighted in \textcolor{red}{red}. 

\begin{table}[htbp]
\begin{tabular}{c | cccc cccc}
 state\textbackslash $m_j$  & -7/2 & -5/2 & -3/2 & -1/2 & 1/2 & 3/2 & 5/2 & 7/2 \\ \hline
0- &  -0.9675    & 0    & 0    & 0.2179    & 0   & 0   & +0.1279   & 0   \\
0+ & 0    &  -0.1279    & 0    & 0    & +0.2179   & 0   & 0   & +0.9675   \\
1- & 0    &  -0.9723    & 0    & 0    &  0.1649   & 0   & 0   &  -0.1656   \\
1+ &  -0.1656    & 0    & 0    &  -0.1649    & 0   & 0   &  -0.9723   & 0   \\
2- & 0    & 0    &  -0.9114    & 0    & 0   &  +0.4115   & 0   & 0   \\
2+ & 0    & 0    &  +0.4115    & 0    & 0   &  0.9114   & 0   & 0   \\
3- & 0.1908    & 0    & 0    &  0.9619    & 0   & 0   &  -0.1957  & 0   \\
3+ & 0    &  -0.1957    & 0    & 0    &  -0.9619   & 0   & 0   & 0.1908  
\end{tabular}
\caption{\label{tab:table-cs-eigen}Eigenstates for CsYbSe$_2$}
\end{table}

\begin{table}[htbp]
\begin{tabular}{c | cccc}
combination\textbackslash $O_{op}$ & $J_+J_+$ & $J_-J_-$ & $J_+J_-$ & $J_-J_+$ \\\hline
 $|| \langle 1- | O_{op} | 0- \rangle ||^2$  & 0        & 6.55101 & 0   & 0 \\
 $|| \langle 1+ | O_{op} | 0- \rangle ||^2$  & 0        & 0       & \textcolor{red}{4.12602}   & 0.104899 \\
 $|| \langle 1+ | O_{op} | 0+ \rangle ||^2$  & 6.55101  & 0       & 0   & 0 \\
 $|| \langle 1- | O_{op} | 0+ \rangle ||^2$  & 0        & 0       & 0.104899   & \textcolor{red}{4.12602} \\\hline

  $|| \langle 2- | O_{op} | 0- \rangle ||^2$  & 89.6949  & 0       & 0   & 0 \\
 $|| \langle 2+ | O_{op} | 0- \rangle ||^2$  & 0.327506 & 0       & 0   & 0 \\
 $|| \langle 2+ | O_{op} | 0+ \rangle ||^2$  & 0.       & 89.6949 & 0   & 0 \\
 $|| \langle 2- | O_{op} | 0+ \rangle ||^2$  & 0        & 0.327506& 0   & 0  \\\hline
 $|| \langle 3- | O_{op} | 0- \rangle ||^2$  & 0        & 0       & \textcolor{red}{8.08612}         & \textcolor{red}{3.55762
} \\
 $|| \langle 3+ | O_{op} | 0- \rangle ||^2$  & 0        & 4.94038 & 0         & 0       \\
 $|| \langle 3+ | O_{op} | 0+ \rangle ||^2$  & 0        & 0       & \textcolor{red}{3.55762}         & \textcolor{red}{8.08612} \\
 $|| \langle 3- | O_{op} | 0+ \rangle ||^2$  & 4.94038  & 0       & 0         & 0       
\end{tabular}
\caption{\label{tab:cs-sele}Matrix element for  $O_{op}$ $ = J_+J_+$,  $J_-J_-$,  $J_+J_-$,  $J_-J_+$ between CEF eigenstates for CsYbSe$_2$. }
\end{table}

\clearpage
\bibliographystyle{naturemag}
\bibliography{suppreferences}